\documentclass[11pt,a4paper,final]{article}
\usepackage[utf8]{inputenc}
\usepackage{amsmath}
\usepackage{amsfonts}
\usepackage{amssymb}
\usepackage{graphicx}
\usepackage[left=2cm,right=2cm,top=2cm,bottom=2cm]{geometry}
\usepackage{url}
\usepackage{subcaption}
\usepackage{tabularx}
\usepackage{natbib}
\usepackage{color}

\newcommand{\PR}{\mathbb{P}}

\newcommand{\vb}{\mathbf{b}}

\newcommand{\vo}{\mathbf{o}}

\newcommand{\hy}{\hat{y}}

\newcommand{\cE}{\mathcal{E}}

\newcommand{\hh}[1]{\textcolor{black}{#1}}

\begin{document}
\title{Mathematical Notions vs. Human Perception of Fairness:\\ A Descriptive Approach to Fairness for Machine Learning}

\author{
	\makebox[.3\textwidth]{Megha Srivastava}\\
   Stanford University\\
   \url{meghas@stanford.edu} \\
   \and
   \makebox[.3\textwidth]{Hoda Heidari}\\
   ETH Z{\"u}rich\\
   \url{hheidari@inf.ethz.ch} \\
   \and
   \makebox[.3\textwidth]{Andreas Krause}\\
   ETH Z{\"u}rich\\
   \url{krausea@ethz.ch} \\
}

\date{}

\maketitle

\begin{abstract}
Fairness for Machine Learning has received considerable attention, recently. Various mathematical formulations of fairness have been proposed, and it has been shown that it is impossible to satisfy all of them simultaneously. The literature so far has dealt with these impossibility results by quantifying the tradeoffs between different formulations of fairness. Our work takes a different perspective on this issue. Rather than requiring all notions of fairness to (partially) hold at the same time, we ask which one of them is the most appropriate given the societal domain in which the decision-making model is to be deployed. We take a descriptive approach and set out to identify the notion of fairness that best captures \emph{lay people's perception of fairness}. We run adaptive experiments designed to pinpoint the most compatible notion of fairness with each participant's choices through a small number of tests. Perhaps surprisingly, we find that the most simplistic mathematical definition of fairness---namely, demographic parity---most closely matches people's idea of fairness in two distinct application scenarios. This conclusion remains intact even when we explicitly tell the participants about the alternative, more complicated definitions of fairness, and we reduce the cognitive burden of evaluating those notions for them. Our findings have important implications for the Fair ML literature and the discourse on formalizing algorithmic fairness. 
\end{abstract}

\section{Introduction}
%Background
Machine Learning tools are increasingly employed to make consequential decisions for human subjects, in areas such as credit lending~\citep{whitecase}, policing~\citep{policing}, criminal justice~\citep{sentencing}, and medicine~\citep{deo2015machine}. Decisions made by these algorithms can have a long-lasting impact on people's lives and may affect certain individuals or social groups negatively~\citep{sweeney2013discrimination,propublica}. This realization has recently spawned an active area of research into quantifying and guaranteeing fairness for machine learning~\citep{dwork2012fairness,kleinberg2016inherent,hardt2016equality}. 

% Literature
Despite the recent surge of interest in Fair ML, there is no consensus on a precise definition of (un)fairness. 
Numerous mathematical definitions of fairness have been proposed; examples include demographic parity~\citep{dwork2012fairness}, disparate impact~\citep{zafar2017dmt}, equality of odds~\citep{hardt2016equality,feldman2015certifying}, and calibration~\citep{kleinberg2016inherent}. While each of these notions is appealing on its own, it has been shown that they are incompatible with one another and cannot hold simultaneously~\citep{kleinberg2016inherent,chouldechova2017fair}. The literature so far has dealt with these impossibility results by attempting to quantify the tradeoffs between different formulations of fairness, hoping that practitioners will be better positioned to determine the extent to which the violation of each fairness criterion can be socially tolerated (see, e.g.,~\citep{corbett2017algorithmic}). 

% Why a descriptive approach
Our work takes a different perspective on these impossibility results. We posit that fairness is a highly context-dependent ideal and depending on the societal domain in which the decision-making model is deployed, one mathematical notion of fairness may be considered ethically more desirable than other alternatives. So rather than requiring all notions of fairness to (partially) hold at the same time, we set out to determine the most suitable notion of fairness for the particular domain at hand.  
Because algorithmic predictions ultimately impact people's lives, we argue that the most appropriate notion of algorithmic fairness is the one that reflects people's idea of fairness in the given context.
We, therefore, take a \emph{descriptive ethics} approach to \textbf{identify the mathematical notion of fairness that most closely matches lay people's perception of fairness}.

% Methodology
%Throughout we focus on the following \emph{statistical} notions of fairness: statistical parity, equality of odds, predictive value parity, and equality of accuracy. 
Our primary goal is to test the following hypotheses:\footnote{All hypotheses were defined and registered with our institution in advance of the experiments.}
\begin{itemize}
\item \textbf{H1:} In the context of recidivism risk assessment, the majority of subjects' responses is compatible with equality of \emph{false negative/positive rates} across demographic groups.
%\item \textbf{H2:} In the context of loan applications, the majority of subjects' responses will be compatible with equality of false negative rates across protected groups (white vs. African American defendants).
%\item \textbf{H3:} In the context of predictive policing, the majority of subjects' responses will be compatible with equality of false omission and discovery rates across protected groups (white vs. African American defendants).
\item \textbf{H2:} In the context of medical predictions, the majority of subjects' responses is compatible with \emph{equality of accuracy} across demographic groups.
%\item \textbf{H5:} In the context of employment/education, the majority of subjects' responses will be compatible with statistical parity across protected groups (male vs. female applicants).
\item \textbf{H3:} When the decision-making stakes are high (e.g., when algorithmic predictions affect people's life expectancy) participants are more sensitive to accuracy as opposed to equality.
\end{itemize}
Our first hypothesis is inspired by the recent media coverage of the COMPAS criminal risk assessment tool and its potential bias against African-American defendants~\citep{propublica}.
Our second hypothesis is informed by the growing concern over the fact that medical experiments are largely conducted with white males as subjects and as a result, their conclusions are reliable only for that particular segment of the population~\citep{taylor2005importance}. Such biases can only be amplified by Machine Learning tools~\citep{medicine}.
Our third hypothesis explores the tension between inequality aversion and accuracy. %Human are notoriously resistant to unequal outcomes, and they are willing to forego some gain to maintain equality. \textbf{H3} posits that 

\begin{figure*}[h!]
	\centering
	\includegraphics[width=.98\textwidth]{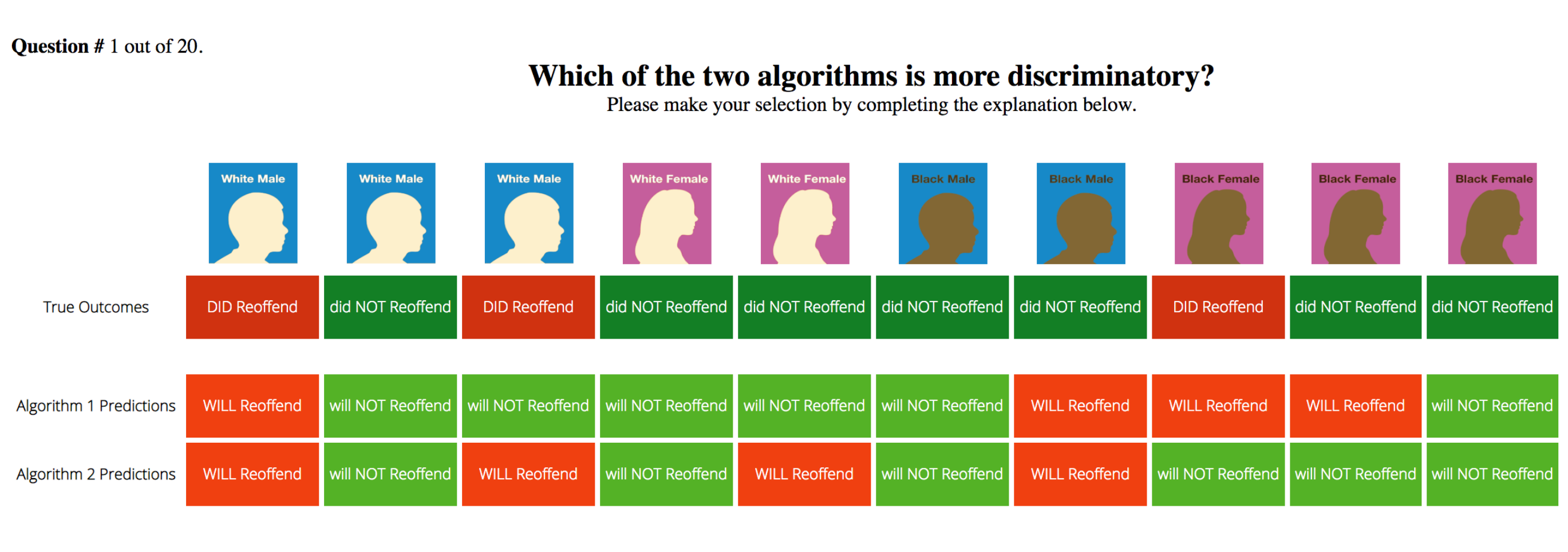}
	\caption{ A typical test in our experiments. Each test illustrates the predictions made by two hypothetical algorithms ($A_1$ and $A_2$) along with the true labels for ten hypothetical individual decision subjects. Decision subjects' race and gender are color-coded. A pink/blue background specifies a female/male decision-subject. A beige/brown contour specifies a Caucasian/African- American decision subject. The demographic characteristics of decision subjects are kept constant across all tests. With this information, the participant responds to the test by selecting the algorithm he/she considers to be discriminatory.}
	\label{fig:test}
\end{figure*}

% Design of experiments
In order to determine the notion of fairness that is most compatible with a participant's perception of fairness, we design an adaptive experiment based on active learning. 
Each participant is required to answer a series of at most 20 adaptively chosen tests---all concerning a fixed, carefully specified scenario (e.g., predicting the risk of future crime for defendants). Each question specifies the ground truth labels for ten hypothetical decision subjects, along with the labels predicted for them by two hypothetical predictive models./algorithms (see Figure~\ref{fig:test}). The participant is then prompted to choose which of the two algorithms they consider to be more discriminatory. The tests are chosen by an active learning algorithm, called EC$^2$~\citep{golovin2010near,ray2012bayesian}. Depending on the participant's choices so far, the EC$^2$ algorithm chooses the next test. Note that an active learning scheme is necessary for pinpointing the most compatible notion through a small number of tests. Without an adaptive design, participants would have to answer hundreds of questions before we can confidently specify the notion of fairness that is most compatible with their choices.

% Empirical findings and insights
We run our experiments on the Amazon Mechanical Turk (AMT) platform and report the percentage of participants whose choices match each mathematical notion of fairness. We also investigate how this varies from one context to another and from one demographic group to another.
Surprisingly, we find that the most simplistic mathematical definition of fairness---namely, \textit{demographic parity}---most closely matches people's idea of fairness in two distinct scenarios. This remains the case even when subjects are explicitly told about the alternative, more complicated definitions of fairness and the cognitive burden of evaluating those notions is reduced for them. 
Our findings have important implications for the Fair ML literature. In particular, they accentuate the need for a more comprehensive understanding of the human attitude toward fairness for predictive models. %Algorithmic decisions will ultimately impact human subjects' lives, and it is, therefore, critical to involve them in the process of choosing the right notion of fairness. 

% Concluding remarks, broader context
In summary, we provide a comparative answer to the ethical question of ``what is the most appropriate notion of fairness to guarantee in a given societal context?''. 
 Our framework can be readily adapted to scenarios beyond the ones studied here. %Our publicly available website can serve as an educational tool to familiarize everyday people with various aspects of fairness for predictive models. 
 As the Fair ML literature continues to grow, we believe it is critical for all stakeholders---including people who are potential subjects of algorithmic decision-making in their daily lives---to be heard and involved in the process of formulating fairness. Our work takes an initial step toward this agenda. After all, a theory of algorithmic fairness can only have a meaningful positive impact on society if it reflects people's sense of justice.
%We envision our work to have a significant contribution to the ongoing debate on formal definitions of fairness, by providing a comparative answer to the critical question of ``what is the best notion of fairness to guarantee in a specific context?''. 

\subsection{Related Work}\label{sec:related}
Much of the existing work on algorithmic fairness has been devoted to the study of \emph{discrimination} (also called \emph{statistical}- or \emph{group}-level fairness) for \emph{binary classification}. Statistical notions require a particular metric---quantifying \emph{benefit} or harm---to be equal across different social groups (e.g., gender or racial groups). Different choices for the benefit metric have led to different fairness criteria; examples include Demographic Parity (DP)~\citep{dwork2012fairness}, Error Parity (EP)~\citep{buolamwini2018gender}, equality of False Positive or False Negative rates (FPP and FNP, respectively)~\citep{zafar2017dmt,hardt2016equality}, and False Discovery or Omission rates (FDP and FOP, respectively)~\citep{zafar2017dmt,kleinberg2016inherent}. 
Demographic parity seeks to equalize the percentage of people who are predicted to be positive across different groups. Equality of false positive/negative rates requires the percentage of people falsely predicted to be positive/negative to be the same across true negative/positive individuals belonging to each group. Equality of false discovery/omission rates seeks to equalize the percentage of false positive/negative predictions among individuals predicted to be positive/negative in each group.
See Table~\ref{tab:benefit} for the precise definition of benefit corresponding to each notion of these notions of fairness. 

\begin{table}[h]
\caption{The measure of benefit/harm corresponding to each notion of fairness. For a decision subject $i$ belonging to group $G$, $y_i$ specifies his/her true label, and $\hy_i$ his/her predicted label. $n_G$ is the number of individuals belonging to group $G$.}
\label{tab:benefit}
\centering
\begin{tabular}{l l l l l l}
Fairness notion & benefit for group $G$ \\ \hline \hline
DP  &  $b^G = \frac{1}{n_G}\sum_{i \in G} 1[\hy_i = 1]$ \\ \hline
EP  &  $b^G = \frac{1}{n_G}\sum_{i \in G} 1[\hy_i \neq y_i]$ \\ \hline
FDP &  $b^G = \frac{\sum_{i \in G} 1[y_i = 0 \& \hy_i = 1]}{\sum_{i \in G} 1[\hy_i = 1]}$ \\ \hline
FNP  &  $b^G = \frac{\sum_{i \in G} 1[\hy_i = 0 \& y_i = 1]}{\sum_{i \in G} 1[y_i = 1]}$ \\ \hline
\end{tabular}
\end{table}

Following~\citet{speicher2018a}, we extend these existing notions of fairness to measures of unfairness by employing inequality indices, and in particular the Generalized Entropy index. Given the benefit vector, $\vb$, computed for all groups $G=1,\cdots, N$, the Generalized Entropy (with exponent $\alpha=2$) is calculated as follows:
\begin{equation}\label{eq:GE}
\cE(\vb) = \cE(b_1,b_2,\cdots,b_N) = \frac{1}{2n} \sum_{G=1}^N \left[ \left(\frac{b_G}{\mu}\right)^2 -1\right].
\end{equation}
where $\mu = \frac{1}{N}\sum_{G=1}^N b_G$ is the average benefit across all groups.

In ethics, there are two distinct ways of addressing moral dilemmas: \emph{descriptive} vs. \emph{normative} approach. 
Normative ethics involves creating or evaluating moral standards to decide what people \emph{should do} or whether their current moral behavior is reasonable. 
Descriptive (or comparative) ethics is a form of empirical research into the attitudes of individuals or groups of people towards morality and moral decision-making. Our work belongs to the descriptive category. Several prior papers have taken a normative perspective on algorithmic fairness. For instance, \citet{gajane2017formalizing} attempt to cast algorithmic notions of fairness as instances of existing theories of justice. \citet{heidari2019a} propose a framework for evaluating the assumptions underlying different notions of fairness by casting them as special cases of economic models of equality of opportunity. We emphasize that there is no simple, widely-accepted, normative principle to settle the ethical problem of algorithmic fairness. 

Several recent papers empirically investigate the issues of fairness and interpretability utilizing human-subject experiments. Below we elaborate on some of the prior work in this line.
MIT's moral machine~\citep{awad2018moral} provides a crowd-sourcing platform for aggregating human opinion on how self-driving cars should make decisions when faced with moral dilemmas. For the same setting, \citet{noothigattu2018voting} propose learning a random utility model of individual preferences, then efficiently aggregating those individual preferences through a social choice function.  \citep{lee2018webuildai} proposes a similar approach for general ethical decision-making. Similar to these papers, we obtain input from human-participants by asking them to compare two alternatives from a moral standpoint. \citeauthor{noothigattu2018voting} and \citeauthor{lee2018webuildai} focus on modeling human preferences and \emph{aggregating} them utilizing tools from \emph{social choice theory}. In contrast, our primary goal is to understand the relationship between human perception and the recently proposed mathematical formulations of fairness. 

\citet{grgic2018human} study why people perceive the use of certain features as unfair in making a prediction about individuals.
\citet{binns2018s} study people's perceptions of justice in algorithmic decision-making under different \emph{explanation styles} and at a high level show that there may be no ``best" approach to explaining algorithmic decisions to people. %\citet{poursabzi2018manipulating} present an experimental framework for assessing the effects of model interpretability on users' \emph{trust} in algorithmic predictions. %, as a function the number of features and the model transparency (clear or black-box).
%\citet{dressel2018accuracy} show that the widely used commercial risk assessment software COMPAS is no more accurate or fair than \emph{predictions made by people} (AMT users) with little or no criminal justice expertise. %\citep{tan2018investigating}
%
\citet{veale2018fairness} interview \emph{public sector machine learning practitioners} regarding the challenges of incorporating public values into their work.
\citet{holstein2018improving} conduct a systematic investigation of \emph{commercial product team}s' challenges and needs  in developing fairer ML systems through semi-structured interviews. Unlike our work where the primary focus is on lay people and potential decision subjects, \citeauthor{holstein2018improving} and \citeauthor{veale2018fairness} study ML practitioners' views toward fairness. 

Several recent papers in human-computer interaction study users' expectations and perceptions related to fairness of algorithms.
\citet{lee2017algorithmic} investigate people's perceptions of \emph{fair division} algorithms (e.g., those designed to divide rent among tenants) compared to discussion-based group decision-making methods.
\citet{woodruff2018qualitative} conduct workshops and interviews with participants belonging to certain marginalized groups (by race or class) in the US to understand their reactions to algorithmic unfairness.

To our knowledge, no prior work has conducted experiments with the goal of mapping existing \emph{group} definitions of fairness to human perception of justice. 
\citet{saxena2018fairness} investigate ordinary people's attitude toward three notions of \emph{individual fairness} in the context of \emph{loan decisions}. They investigate the following three notions: 1) treating similar individuals similarly~\citep{dwork2012fairness}; 2) never favoring a worse individual over a better one~\citep{joseph2016fairness}; 3) the probability of approval being proportional to the chance of the individual representing the best choice. They show that people exhibit a preference for the last fairness definition---which is similar, in essence, to calibration.

\section{Study Design and Methodology}\label{sec:methodology}
To test \textbf{H1} and \textbf{H2}, we conducted human-subject experiments on AMT to determine which one of the existing notions of group fairness best captures people's idea of fairness in the given societal context. For our final hypothesis, \textbf{H3}, we utilized short survey questions (see Section~\ref{sec:survey} for further details).

\subsection{User Interfaces}
In our experiments, each participant was asked to respond to a maximum of 20 tests. In each test, we showed the participant the predictions made by two hypothetical algorithms for ten hypothetical individuals, along with the true labels (see Figure~\ref{fig:test}). We limited attention to tests that consist of algorithms with equal overall accuracy to control for the impact of accuracy on people's perception. Note that the pair of algorithms displayed in each test changes from one test to another.  We asked the participant to specify which algorithm they consider to be more discriminatory. The tests were chosen adaptively to find the notion of fairness most compatible with the participant's choices using only a small number of tests. The participant was then required to provide an \emph{explanation} for their choice.%For each of the 10 individuals, we  show the participant some of their attribute values the algorithm used to make a prediction about the individual. 

\begin{figure}[h!]
	\centering
%	\begin{subfigure}[b]{0.48\textwidth}
%		\includegraphics[width=\textwidth]{Figures/explanation2.png}
%		\caption{}
%		\label{fig:explanation2}
%	\end{subfigure}
%	\begin{subfigure}[b]{0.48\textwidth}
		\includegraphics[width=0.48\textwidth]{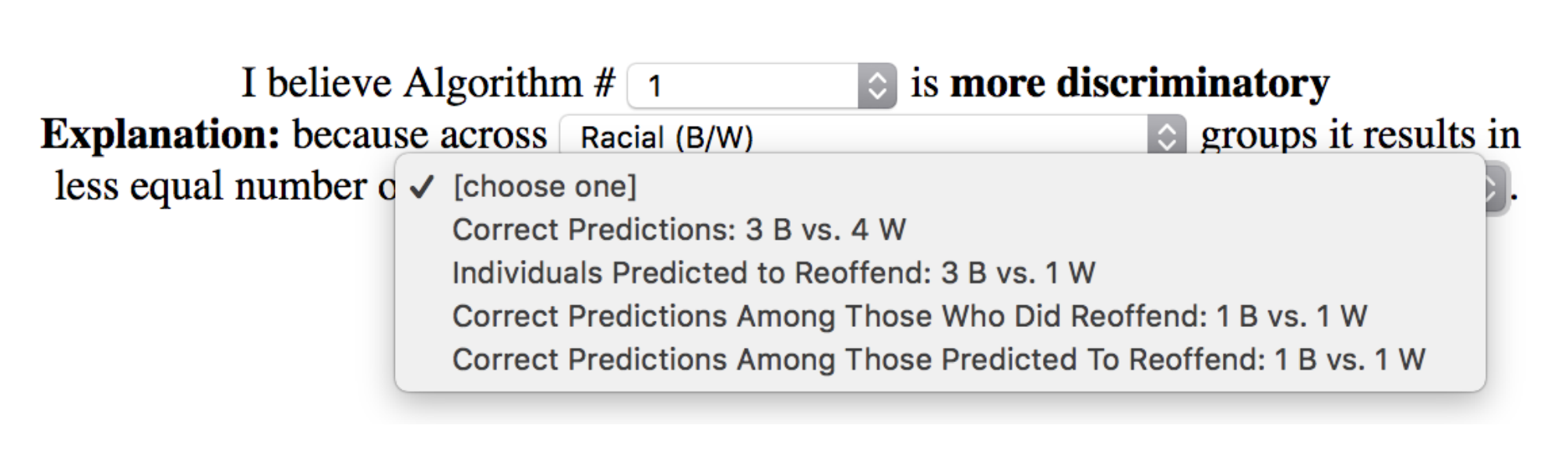}
%		\caption{}
%		\label{fig:dropdown2}
%	\end{subfigure}
	\caption{ The user interface eliciting structured explanations from participants. All benefit metrics are computed and displayed to reduce the cognitive burden of evaluating our fairness notions.}\label{fig:UI2}
\end{figure}

To elicit the explanations, we designed and tested two different User Interfaces (UI). The first UI presents the participant with a text box, allowing them to provide an unstructured explanation/reasoning for their choice. The second UI requires the participant to provide a structured explanation, and it consists of two drop-down menus. In the first dropdown menu, the participant must choose the demographic characteristic they think the algorithm is most discriminatory with respect to (that is, gender, race, or the intersection of the two). In the second drop-down menu, they choose the metric along which the think the algorithm is most discriminatory. To reduce the cognitive burden, we computed the benefit metric corresponding to each notion of fairness and displayed it in the second drop-down menu. See Figure~\ref{fig:UI2}.

We validated our interface design through two rounds of \emph{pilot} studies---one internal (among members of our research group) and one on AMT (among 20 crowd workers). We made some minor changes after our internal run to improve readability of the task description. The AMT pilot participants found our first user interface (the one with text explanations) less restrictive and easier to work with, so we scaled up our experiments to 100 crowd workers per scenario using that interface.

\subsection{Scenarios and Contexts}
We ran our experiments for two distinct prediction tasks: criminal and skin cancer risk prediction. Below we describe each scenario precisely as it was shown to the study participants.

\paragraph{\textbf{Criminal risk prediction}} \textit{Across the United States, data-driven decision-making algorithms are increasingly employed to predict the likelihood of future crimes by defendants. These algorithmic predictions are utilized by judges to make sentencing decisions for defendants (e.g., setting the bond amount; time to be spent in jail).
Data-driven decision-making algorithms use historical data about past defendants to learn about factors that highly correlate with criminality. For instance, the algorithm may learn from past data that: 1) a defendant with a lengthy criminal history is more likely to reoffend if set free---compared to a first-time defender, or
2) defendants belonging to certain groups (e.g., residents of neighborhoods with high crime rate) are more likely to reoffend if set free. 
However, algorithms are not perfect, and they inevitably make errors----although the error rate is usually very low, the algorithm's decision can have a significant impact on some defendants' lives. A defendant falsely predicted to reoffend can unjustly face longer sentences, while a defendant falsely predicted not to reoffend may commit a crime that was preventable.}

\paragraph{\textbf{Skin cancer risk prediction}} \textit{Data-driven algorithms are increasingly employed to diagnose various medical conditions, such as risk for heart disease or various forms of cancer. They can find patterns and links in medical records that previously required great levels of expertise and time from human doctors. Algorithmic diagnoses are then used by health-care professionals to create personalized treatment plans for patients (e.g., whether the patient should undergo surgery or chemotherapy).
Data-driven decision-making algorithms use historical data about past patients to learn about factors that highly correlate with risk of cancer. For instance, the algorithm may learn from past data that: 1) a patient with a family history of skin cancer has a higher risk of developing skin cancer; or 2) patients belonging to certain groups (e.g., people with a certain skin tone, or people of a certain gender) are more likely to develop skin cancer. 
However, algorithms are not perfect, and they inevitably make errors---although the error rate is usually very low, the algorithm's decision can have a significant impact on patients' lives. 
A patient falsely diagnosed with high risk of cancer may unnecessarily undergo high-risk and costly medical treatments, while a patient falsely labeled as low-risk for cancer may face a lower chance of survival.}

\subsection{Adaptive Experimental Design}\label{sec:adaptive}
\begin{figure}[t!]
	\centering
	\begin{subfigure}[b]{0.48\textwidth}
		\includegraphics[width=\textwidth]{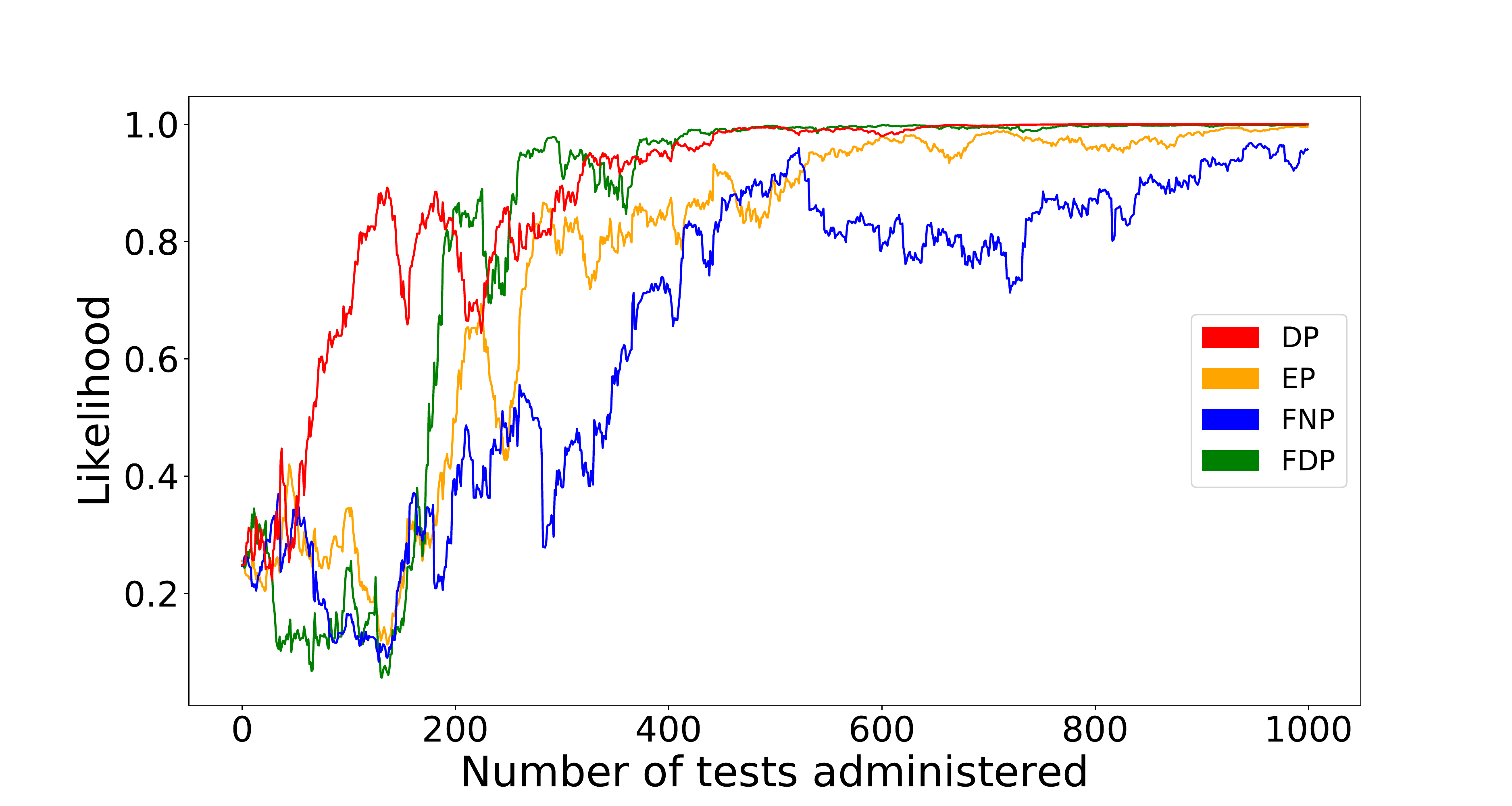}
%		\caption{}
%		\label{fig:test_random}
	\end{subfigure}
	\begin{subfigure}[b]{0.48\textwidth}
		\includegraphics[width=\textwidth]{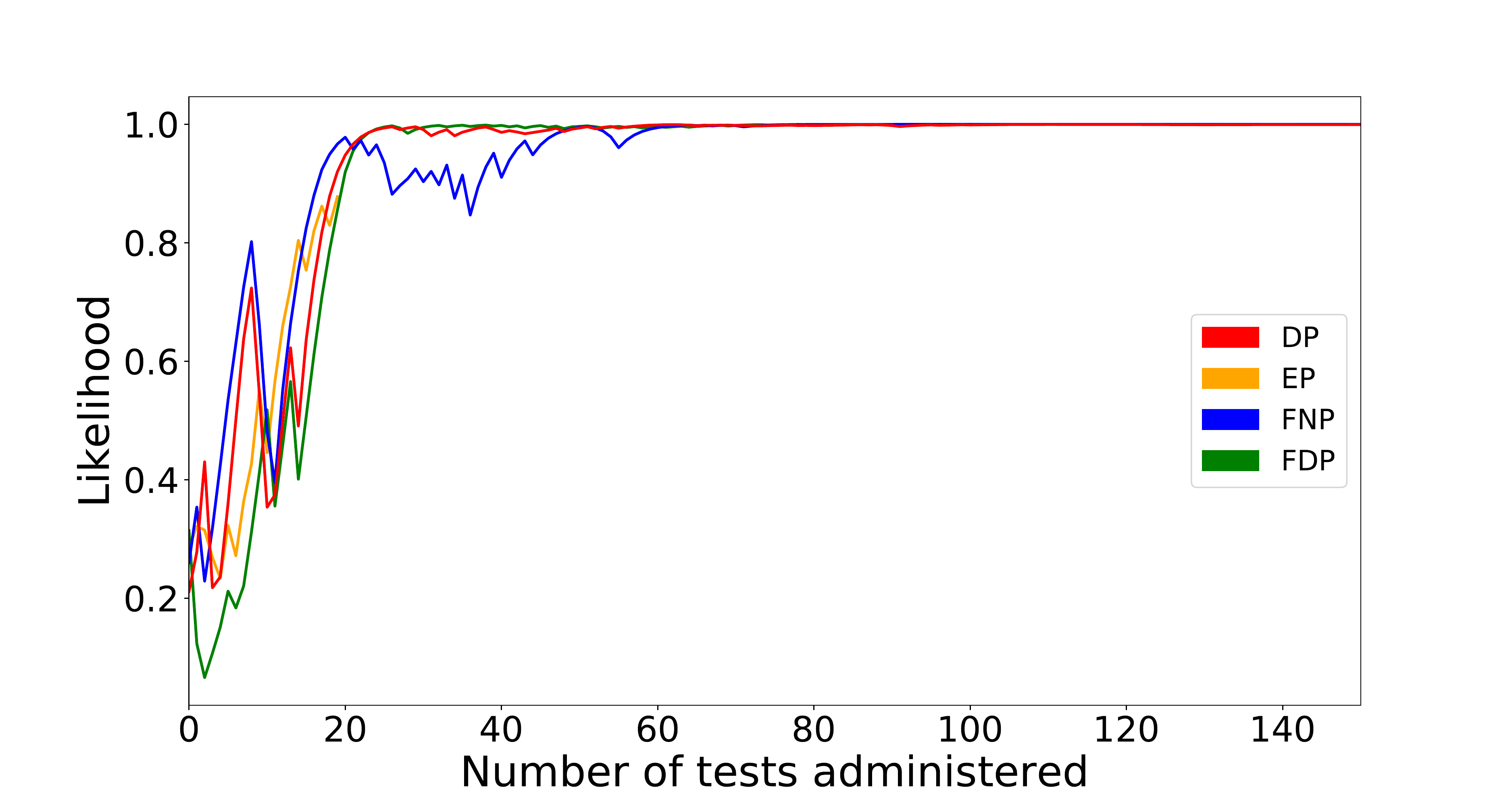}
%		\caption{}
%		\label{fig:test_ec2}
	\end{subfigure}
	\caption{The advantage of adaptive over random test selection. For each of the four notions of fairness, we simulated a participant who follows that notion---as defined by our noisy response model---for 1000 tests. (Top) With a random test sequence, at least 600 tests are needed before we can obtain a high likelihood for one hypothesis. (Bottom) With our adaptive test selection method, only around 20 tests are needed to assign a high likelihood to the fairness notion the participant is following.}\label{fig:ec2_benefit}
\end{figure}

Each run of our experiment consists of sequentially selecting among a set of noisy tests and observing the participant's response to it. These tests are time-consuming for participants to evaluate and they can quickly become repetitive.\footnote{Every participant who completed our task was paid 5 USD. } So to limit the number of tests per participant, we employed an active learning scheme proposed by \citet{golovin2010near}. This allowed us to determine the most compatible fairness notion with each participant's choices, employing at most 20 tests per participant (see Figure~\ref{fig:ec2_benefit}). The algorithm, called EC$^2$ (for Equivalence Class Edge Cutting algorithm), can handle noisy observations and its expected cost is competitive with that of the optimal sequential policy.

\paragraph{\textbf{The EC$^2$ algorithm}}
At a high level, the algorithm works as follows: it designates an \emph{equivalence class} for each notion of fairness. In our setting, we have four equivalence classes representing Demographic Parity (DP), Error Parity (EP), False Discovery rate Parity (FDP), and False Negative rate Parity (FNP). (For  experiments with a different set of hypotheses excluding DP and including other group notions of fairness, such as False Positive Parity (FPP) and False Omission Parity (FOP), see Appendix~\ref{app:additional}.) We use the notation $h_1, h_2, h_3, h_4$ to refer to these notions, respectively. We assume a uniform prior over $h_1,\cdots,h_4$.% are equally probable.

Let $\tau$ denote the number of all possible tests we can run. In our case, $\tau = 9262$. % is upper-bounded by $2^{30}$ (binary choice for each of the ten true labels, ten labels predicted by algorithm 1, and ten  labels predicted by algorithm 2). 
We assume the cost is uniform across all tests.
The outcome of each test is binary (the participant chooses one of the two algorithms as more discriminatory), so there will be $2^\tau$ possible test outcome profiles, where a test outcome profile specifies the outcome of all $\tau$ tests. 
For each test outcome profile, EC$^2$ computes the fairness notion that maximizes the posterior probability of that outcome profile; we call this the MAP fairness notion for the outcome profile. EC$^2$ puts each outcome profile in the equivalence class corresponding to its MAP fairness notion.

EC$^2$ introduces edges between any two outcome profiles that belong to different equivalence classes.
As the algorithm progresses, some edges are cut, and the algorithm terminates when no edge is left. At a high level, in each step, the algorithm picks the test that if run, in expectation removes as much of the remaining edge weights as possible.\footnote{We implemented an efficient approximation of EC$^2$, called the EffECXtive (Efficient Edge Cutting approXimate objective) algorithm.}

\paragraph{\textbf{The Bayesian update}}
Let $\mathcal{T}$ denote the set of all available tests, and $\mathcal{A}$ the set of tests already carried out in the preceding steps. Let $O_t=$ denote the outcome of the test currently administered, denoted by $t$ (i.e., $O_t \in \{ A_1, A_2\}$), and $\vo$, the vector of outcomes of tests in $\mathcal{A}$. Then at each step we select the test $t \in \mathcal{T}-\mathcal{A}$ such that the following objective function is maximized:
{ \small
\begin{equation*}
\triangle(t \vert \vo)=\left[ \sum _{ o \in \{ A_1, A_2\} }{ \PR(O_{ t } = o \vert \vo)\left( \sum _{ i }{ { \PR(h_{ i } \vert \vo, O_{ t }= o) }^{ 2 } }  \right)  }  \right] -\sum _{ i }{ { \PR(h_{ i }\vert \vo)^2 } }
\end{equation*}
}
The objective function for each candidate test is computed as follows:\footnote{The implementation makes use of memorization in order to speed up the computation and achieve a responsive user interface.}
We assume a uniform prior on our four hypotheses, that is, for $h \in \{ h_1 \cdots, h_4\}$, $\PR(h)=1/4$.
We assume conditional independence, that is, $\PR(\vo \vert h) = \prod_{o \in \vo}{ \PR(o \vert h) }$. We have that $\PR(\vo)=\sum _{h }{ { \PR(\vo \vert h) \PR(h) } }$ and $\PR(h \vert \vo)=\frac { \PR(\vo \vert h)\PR(h) }{ \PR(\vo) }$. Moreover,
${ \PR(O_{ t }=o \vert \vo) }=\sum _{ h }{ { \PR(O_{ t }=o \vert h)\PR(h \vert \vo) } }$.
Calculating $\PR(O_{ t } = o \vert h)$ requires modeling participant's fairness assessments, which we discuss next.

\paragraph{\textbf{Modeling participants' fairness assessments}} After observing the participant's response to a test, EC$^2$ needs to update the posterior probability of each notion of fairness. This requires us to specify a model of how the participant responds to each test conditioned on their preferred notion of fairness. 
We assume\footnote{There are numerous alternative assumptions we could have made; we consider ours to be one reasonable choice.} that the probability of an individual following $h_i$ selecting $A_1$ is as follows:
\begin{equation}
\PR\left( A_1 | h_{ i } \right) = \text{softmax}\left( \cE( \vb_{i,1}) , \cE( \vb_{i,2})  \right)
\end{equation}
where $\vb_{i,1}= \langle b_{i,1}^G \rangle_{\text{ group } G}$ and $\vb_{i,2}=\langle b_{i,2}^G \rangle_{\text{ group } G}$ are the group-level benefit vectors for $A_1$ and $A_2$ respectively.  
 $\cE(\vb)$ is the Generalized Entropy index, a measure of inequality, computed over the benefit vector $\vb$. For the benefit vector corresponding to each notion of fairness, see Table~\ref{tab:benefit}.

\begin{figure}[h!]
	\centering
%	\begin{subfigure}[b]{0.48\textwidth}
		\includegraphics[width=0.48\textwidth]{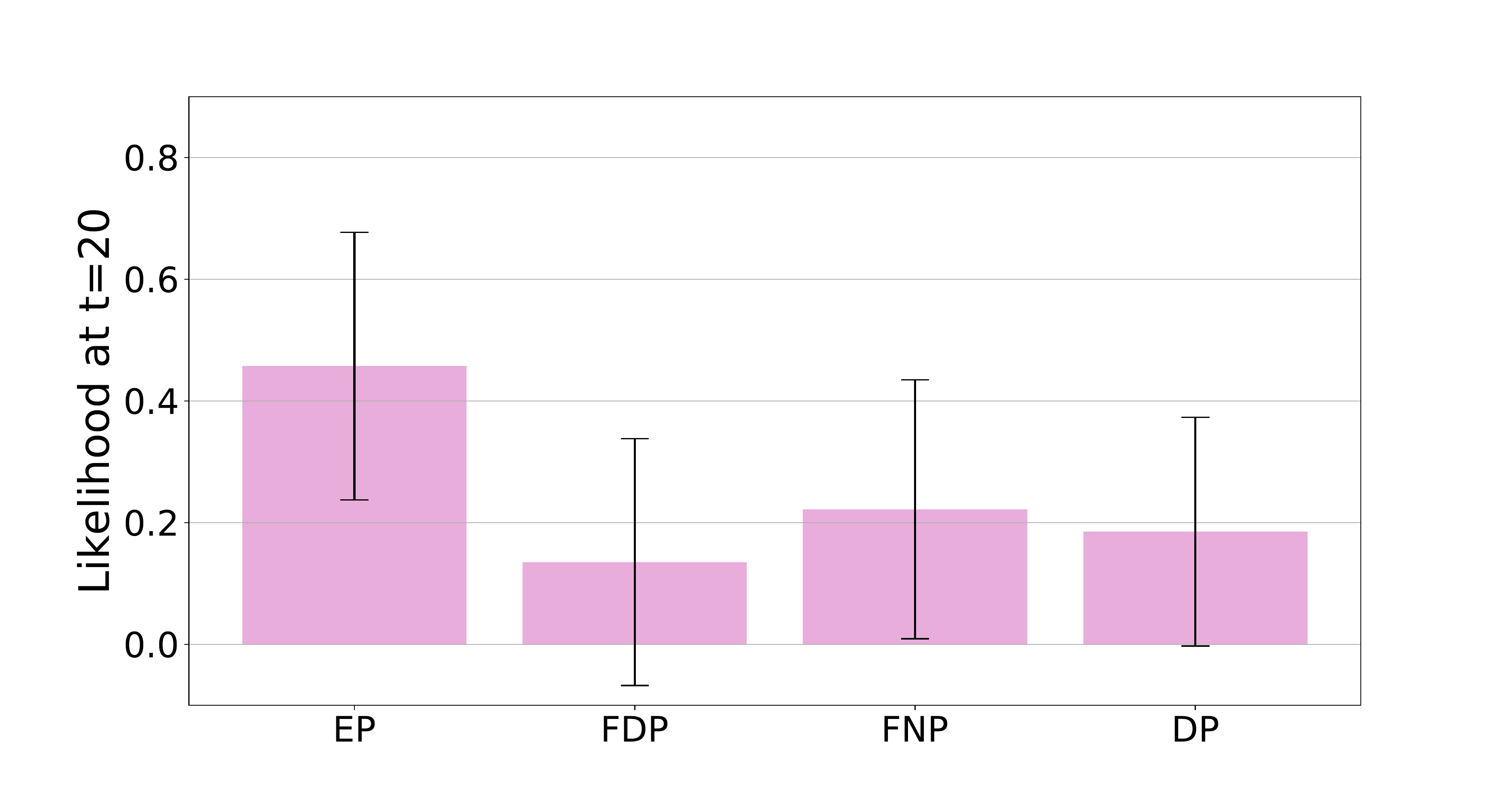}
%		\caption{}
%		\label{fig:baseline_bar}
%	\end{subfigure}
%	\begin{subfigure}[b]{0.48\textwidth}
%		\includegraphics[width=\textwidth]{Figures/Baseline/baseline_confidence.pdf}
%		\caption{}
%		\label{fig:ebaseline_conf}
%	\end{subfigure}
	\caption{The outcome of our adaptive experiment if the response to each test is chosen at random. The chance of DP being selected as the most compatible hypothesis is not high. }\label{fig:baseline}
\end{figure}

Figure~\ref{fig:baseline} shows the outcome of our adaptive experiments, assuming that participants choose their answer to each test uniformly at random. 

\section{Experimental Findings }\label{sec:findings}

%\subsection{Data Set}
Through our experiments on AMT, we gathered a data set consisting of 
\begin{itemize}
\item 100 participants' responses to our tests for the crime risk prediction scenario;
\item 100 participants' responses to our tests for the skin cancer risk prediction scenario.
\end{itemize}
We asked the participants to provide us with their demographic information, such as their age, gender, race, education, and political affiliation. The purpose of asking these questions was to understand whether there are significant variations in perceptions of fairness across different demographic groups.\footnote{Answering to this part was entirely optional and did not affect the participant's eligibility for participation or monetary compensation.}
Table~\ref{tab:demographic} summarizes the demographic information of our participants and contrasts it with the 2016 U.S. census data. In general, AMT workers are not a representative sample of the U.S. population (they often belong to a particular segment of the population---those that have Internet access and are willing to complete crowdsourced tasks online). In particular, for our experiments, participants were younger and more liberal compared to the average U.S. population.

\begin{table}[h]
\caption{Demographics of our AMT participants compared to the 2016 U.S. census data.}
\label{tab:demographic}
\centering
\begin{tabular}{l l l l}
Demographic Attribute & AMT & Census\\ \hline \hline
Male  & 53\% & 49\% \\
Female  & 47\% & 51\% \\\hline
Caucausian  & 68\% & 61\% \\
African-American  & 12\% &  13\%\\
Asian  & 10\% &  6\%\\
Hispanic  & 6\% &  18\%\\ \hline
Liberal  & 74\% &  33\%\\
Conservative  & 19\% &  29\%\\ \hline
High school  & 31\% &  40\%\\
College degree  & 48\% &  48\%\\
Graduate degree  & 20\% &  11\%\\ \hline
18--25  & 14\% & 10\%\\ 
25--40  & 67\% & 20\% \\ 
40--60  & 16\% &  26\%\\ \hline
\end{tabular}
\end{table}

%\begin{figure*}[h!]
%	\centering
%	\includegraphics[width=0.45\textwidth]{Figures/demographics.png}
%	\caption{Participants demographic information.}
%	\label{fig:demographic}
%\end{figure*}

\subsection{Quantitative Results}
\begin{figure*}[t!]
	\centering
	\begin{subfigure}[b]{0.24\textwidth}
		\includegraphics[width=\textwidth]{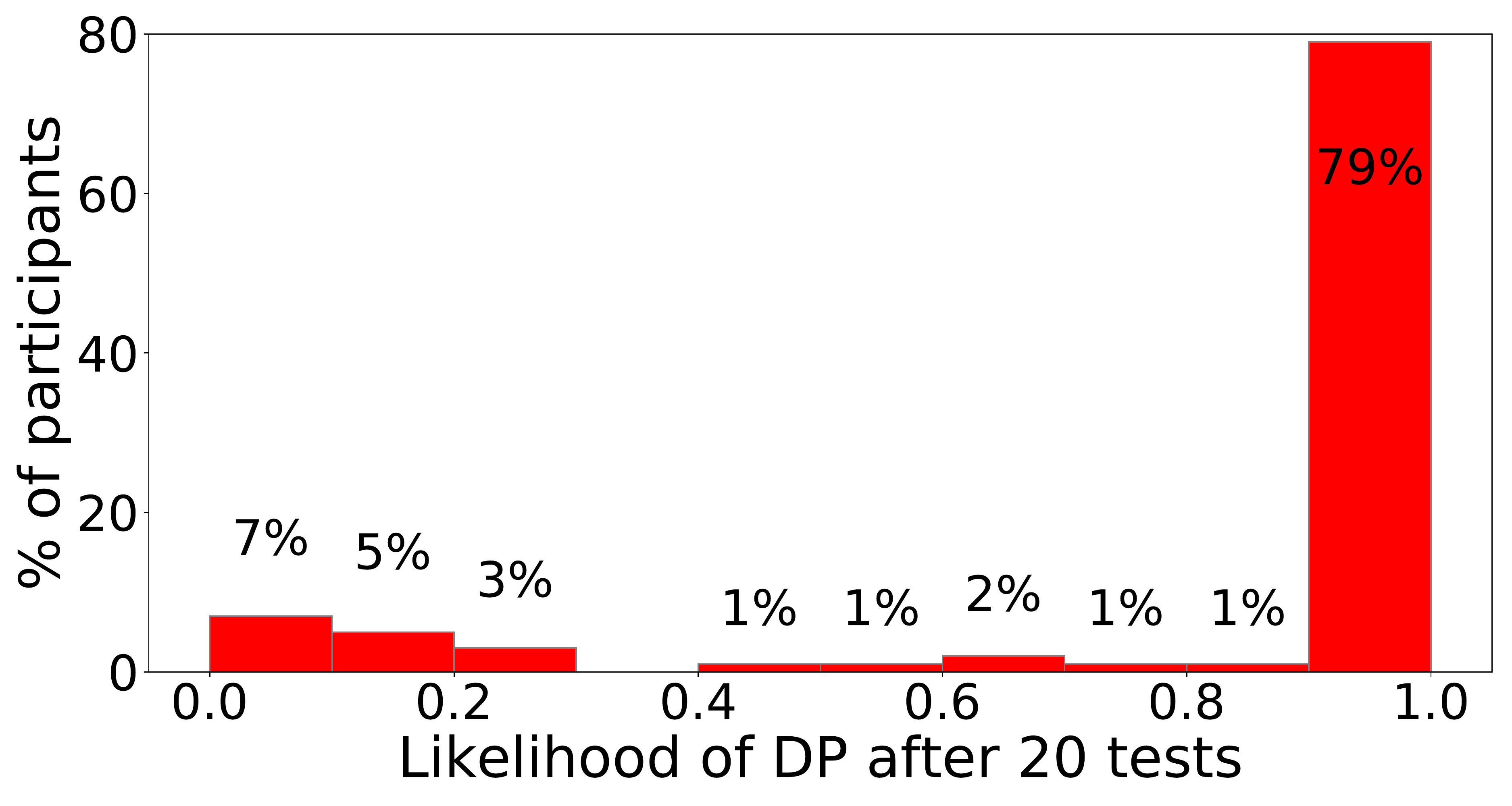}
	\end{subfigure}
	\begin{subfigure}[b]{0.24\textwidth}
		\includegraphics[width=\textwidth]{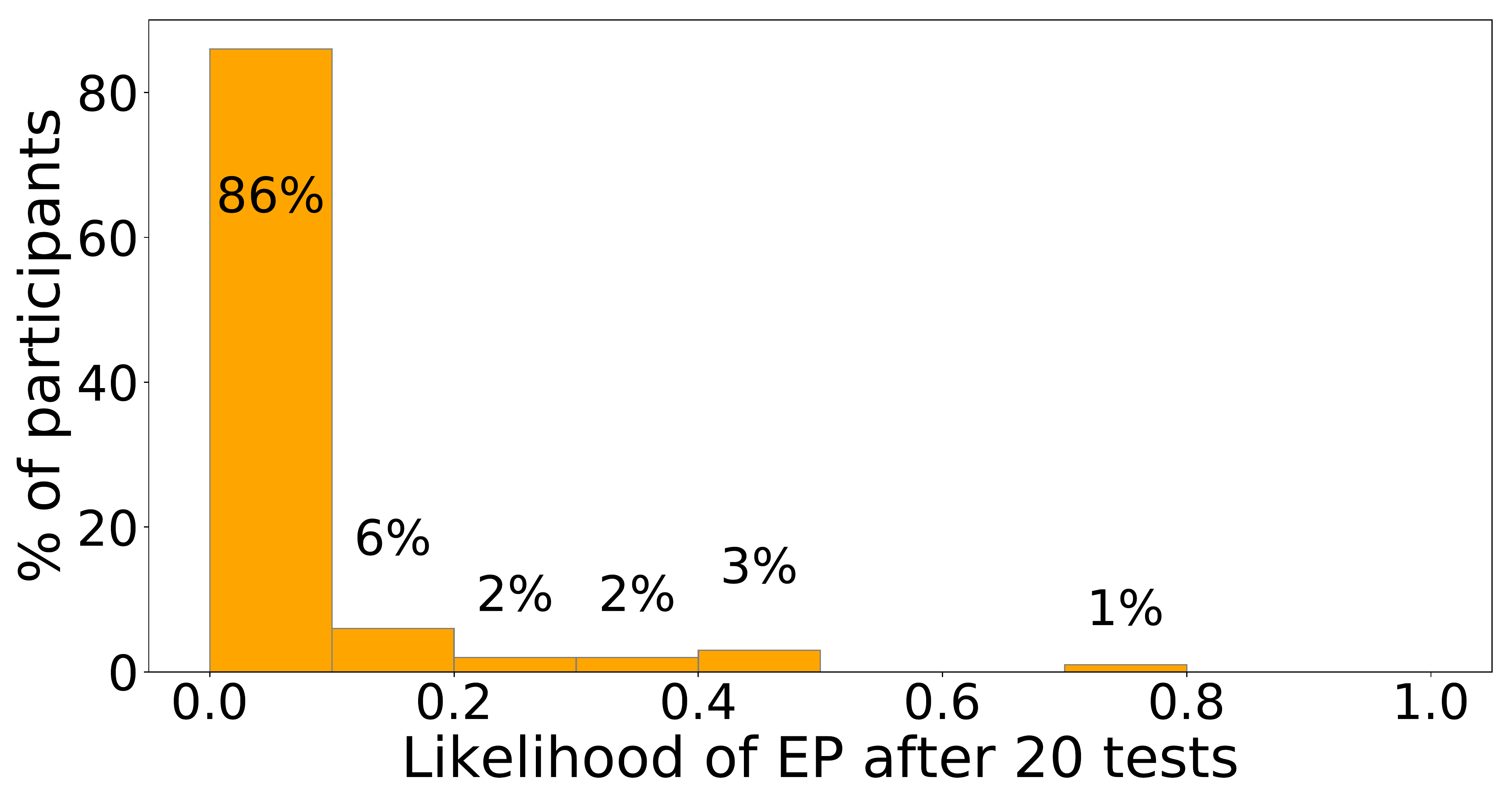}
	\end{subfigure}
	\begin{subfigure}[b]{0.24\textwidth}
		\includegraphics[width=\textwidth]{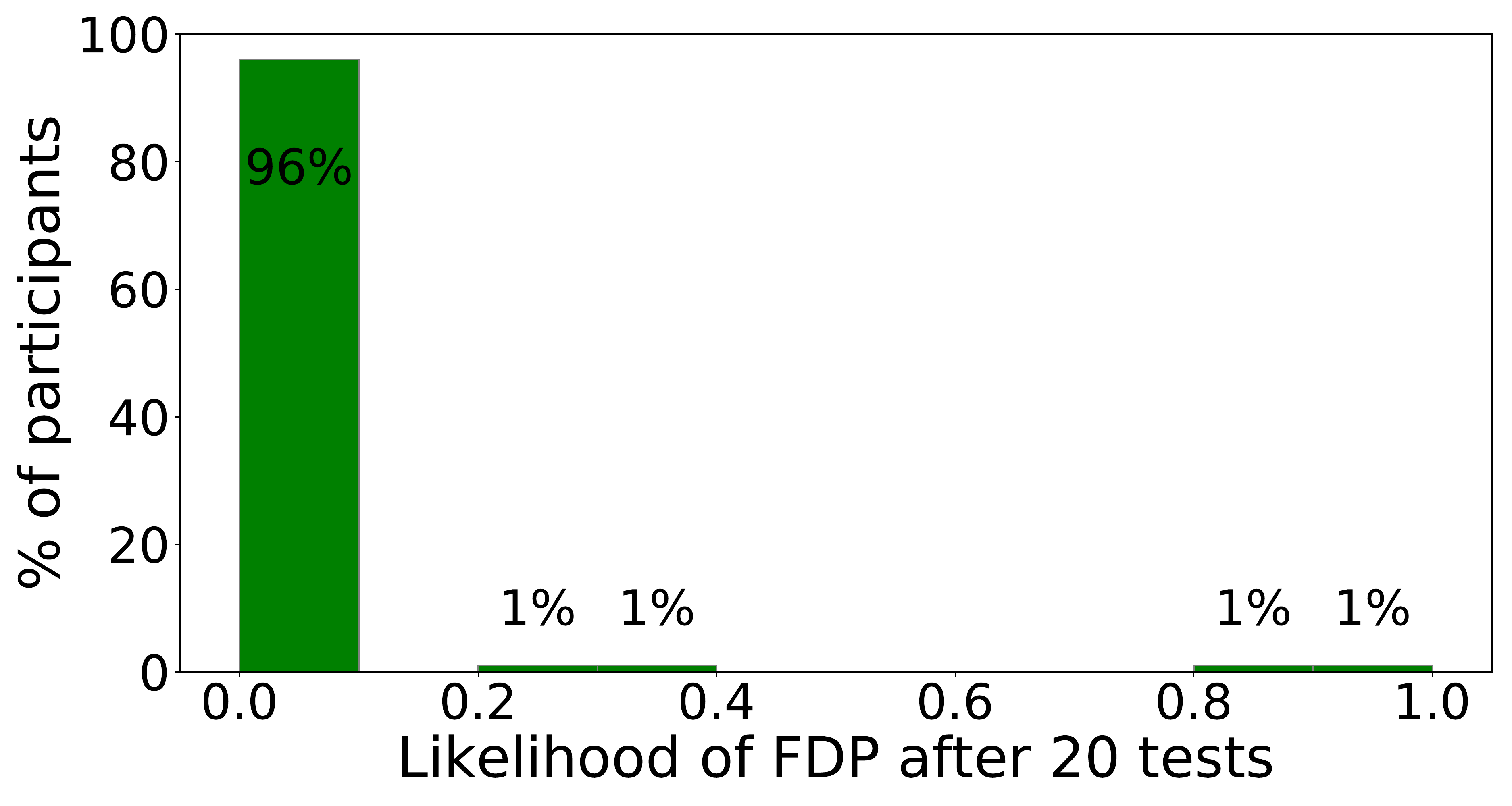}
	\end{subfigure}
	\begin{subfigure}[b]{0.24\textwidth}
		\includegraphics[width=\textwidth]{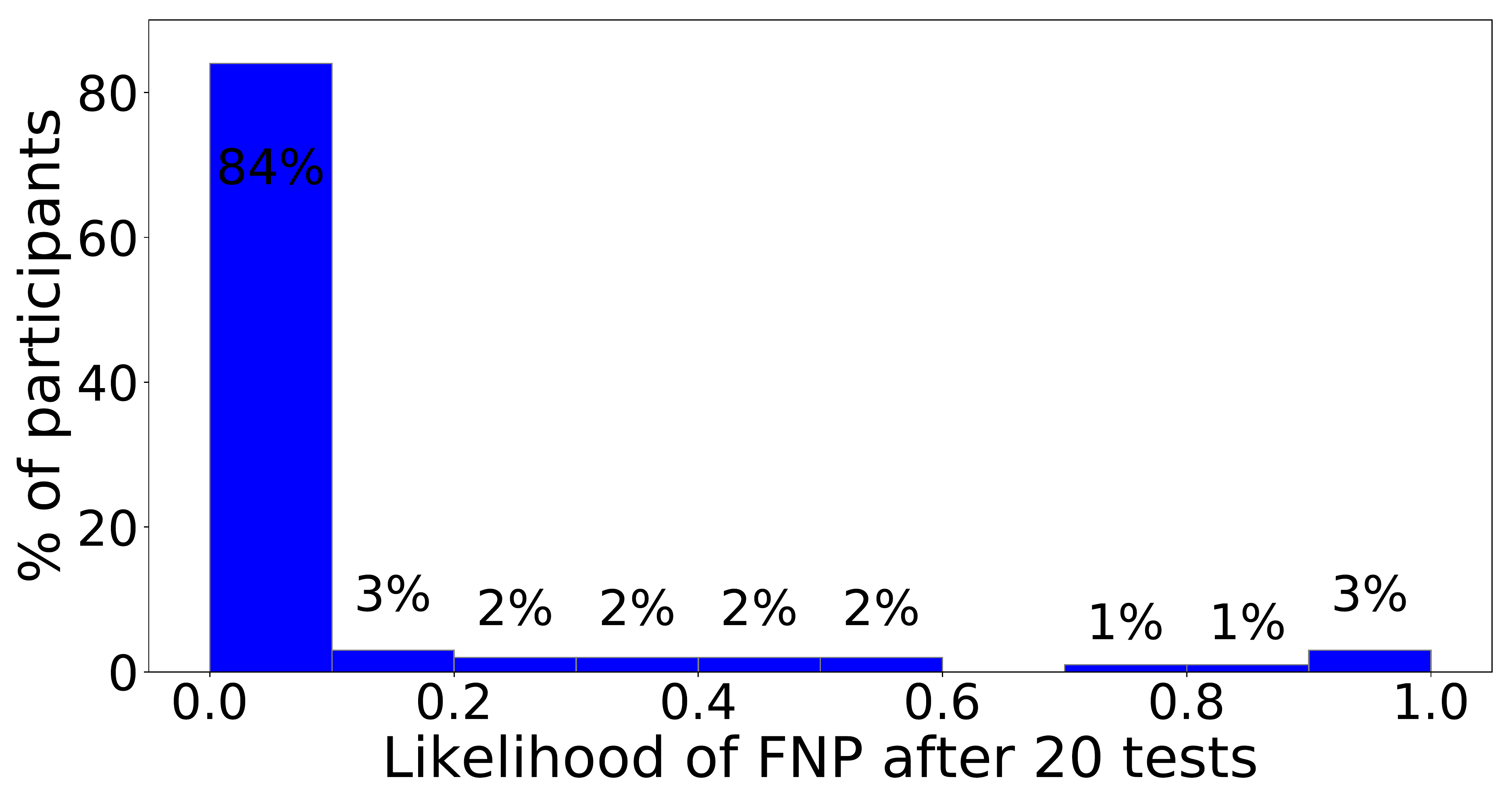}
	\end{subfigure}		
	\caption{The crime risk prediction scenario---the number of participants matched with each notion of fairness (y-axis) along with the likelihood levels (x-axis). Demographic parity captures the choices made by the majority of participants. }\label{fig:histogram_crime}
\end{figure*}

For the crime risk prediction scenario, Figure~\ref{fig:histogram_crime} shows the number of participants whose responses were compatible with each notion of fairness, along with the confidence with which the EC$^2$ algorithm puts them in that category. Demographic parity best captures the choices made by the majority of our participants. Trends are similar for the cancer risk prediction scenario.
See Table~\ref{tab:sum} for the summary. 

\begin{table}[h]
\caption{Number of participants matched with each notion with high likelihood ($>80\%$).}
\label{tab:sum}
\centering
\begin{tabular}{llllll}
  &  DP &  EP &  FDR & FNP & none \\ \hline
Crime risk prediction &  80\% &  0\% &  2\% &  4\% &  14\%\\
Cancer risk prediction & 73\% & 3\%  &  0\% & 0\%  &  24\%\\
\end{tabular}
\end{table}

If we exclude DP from the set of hypotheses (and include other group notions of fairness), the majority of participants can't be confidently categorized as following either one of our hypotheses. This supports our original finding that DP best captures most participants' perception of fairness. See Appendix~\ref{app:additional} for further detail.

\paragraph{\textbf{Sensitivity to explanation interface}} To test the sensitivity of our findings to the design of the user interface, we experimented (at a smaller scale with 20 participants) with an interface that elicits structured explanations from the participant. The interface explicitly shows the disparities across $h_1,\cdots, h_4$. Through this interface, we prompted the participant to think about the fairness notions of interest and reduced the cognitive burden of evaluating those notions for them. We observe that even under this new condition the majority of participants made choices that are best captured by demographic parity.
For the crime prediction scenario, 17 out of 20 participants were matched with DP, and for the cancer prediction scenario, this number was 9 out of 20.

\paragraph{\textbf{Variation across gender, race, age, education, and political views}} We did not observe significant variation across any demographic attribute. For the crime risk prediction scenario, the percentage of subjects whose likelihood of following DP is high ($>80\%$) is as follows: 
\begin{itemize}
\item $78\%$ for female, and $79\%$ for male participants;
\item $82\%$ for Caucasian, and $72\%$ for non-Caucasian participants;
\item $80\%$ for liberal, and $74\%$ for conservative participants;
\item $77\%$ for participants with no college/university degree, and $79\%$ for the rest;
\item $78\%$ for young participants (with age $< 40$), and $81\%$ for older participants.
\end{itemize}

\begin{figure*}[t!]
	\centering
	\begin{subfigure}[b]{0.24\textwidth}
		\includegraphics[width=\textwidth]{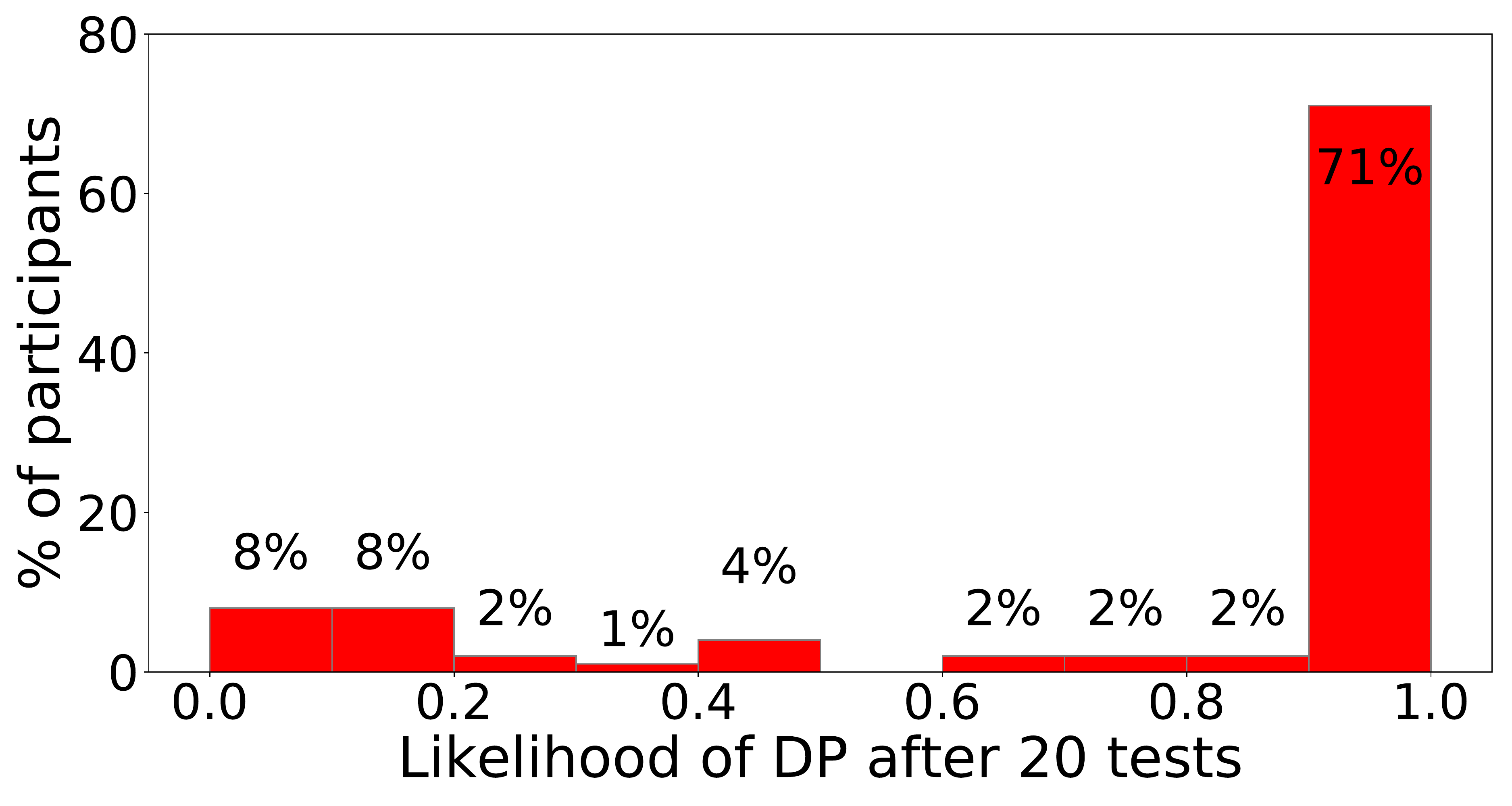}
	\end{subfigure}
	\begin{subfigure}[b]{0.24\textwidth}
		\includegraphics[width=\textwidth]{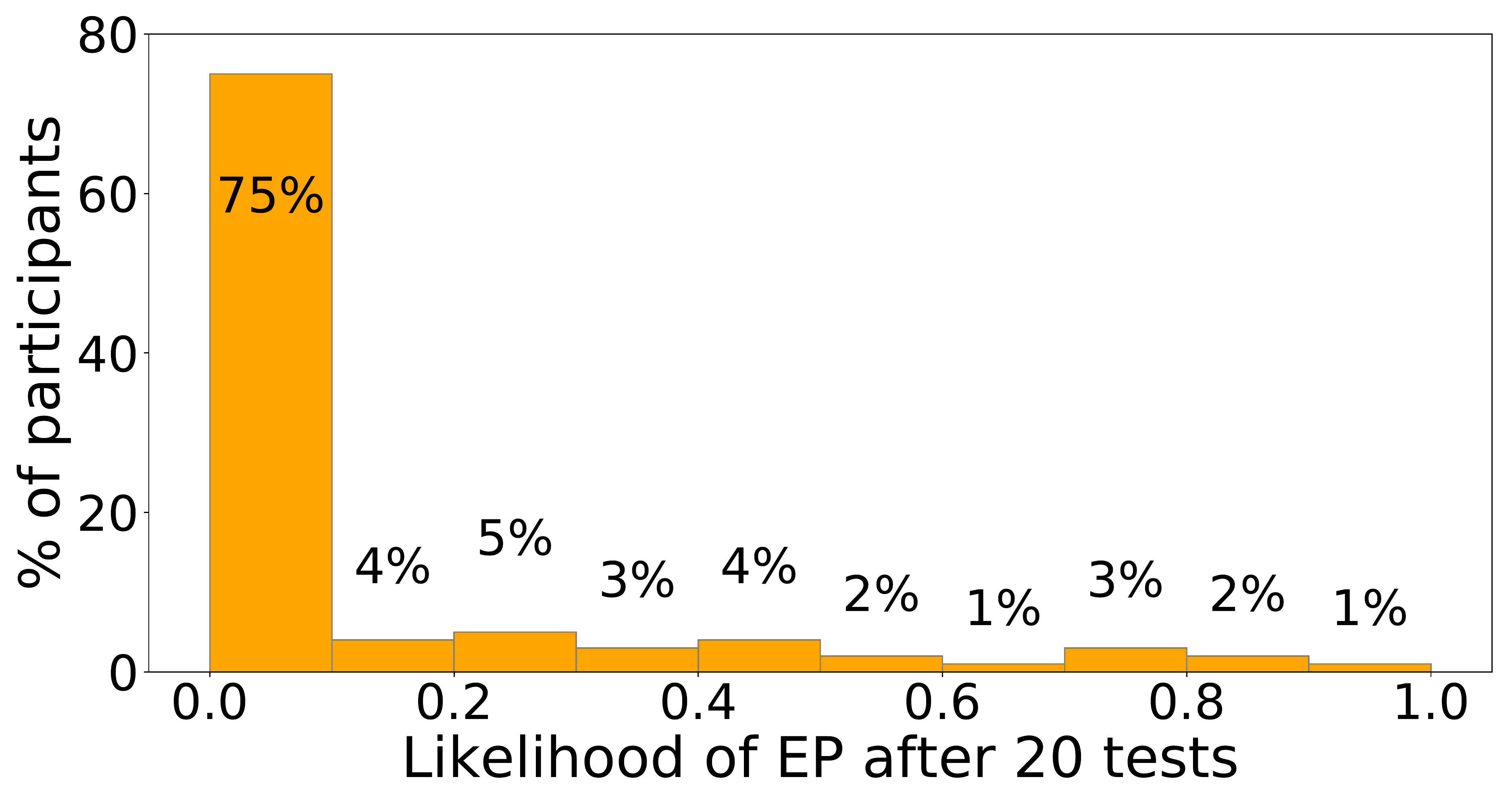}
	\end{subfigure}
	\begin{subfigure}[b]{0.24\textwidth}
		\includegraphics[width=\textwidth]{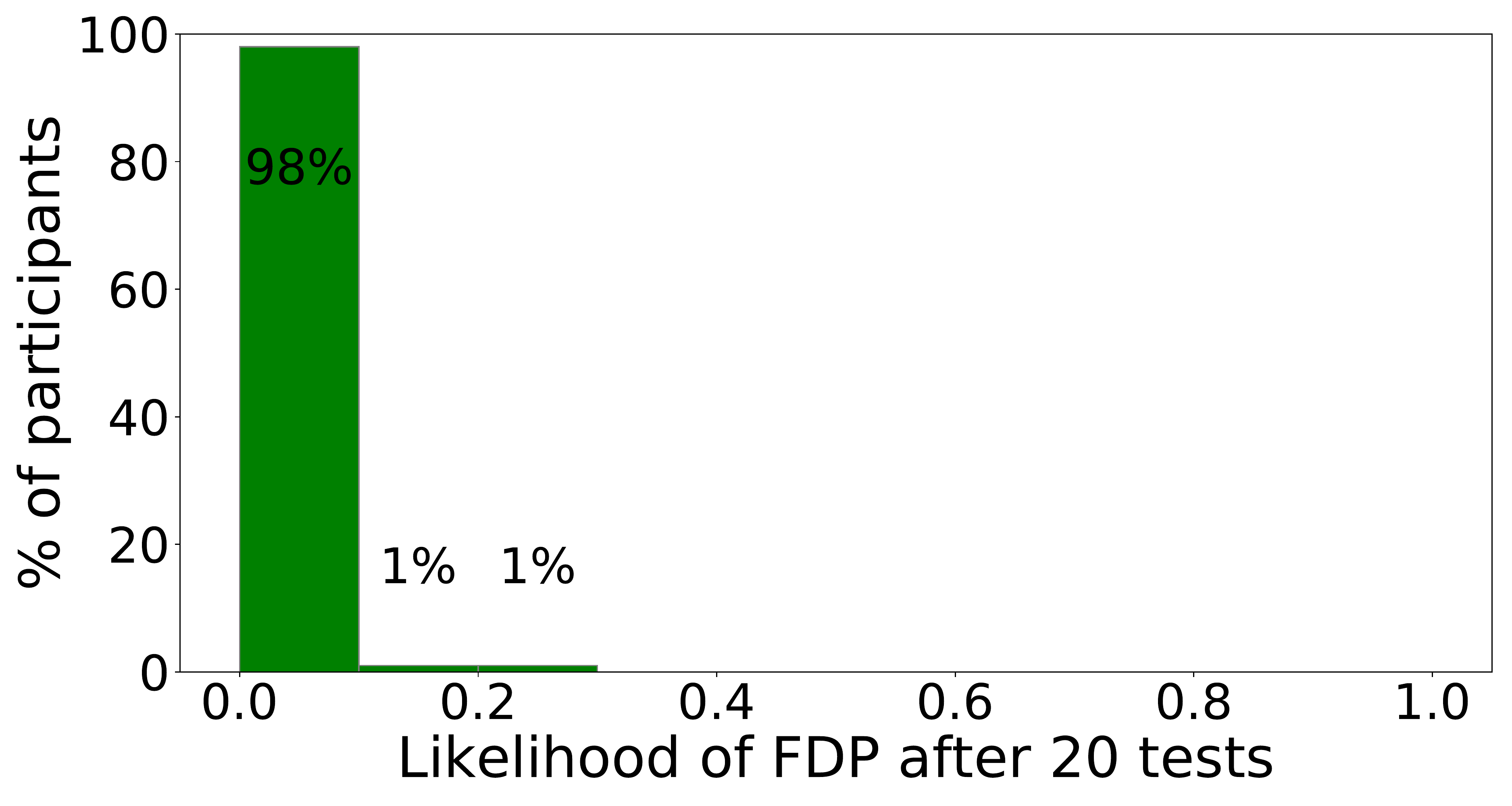}
	\end{subfigure}
	\begin{subfigure}[b]{0.24\textwidth}
		\includegraphics[width=\textwidth]{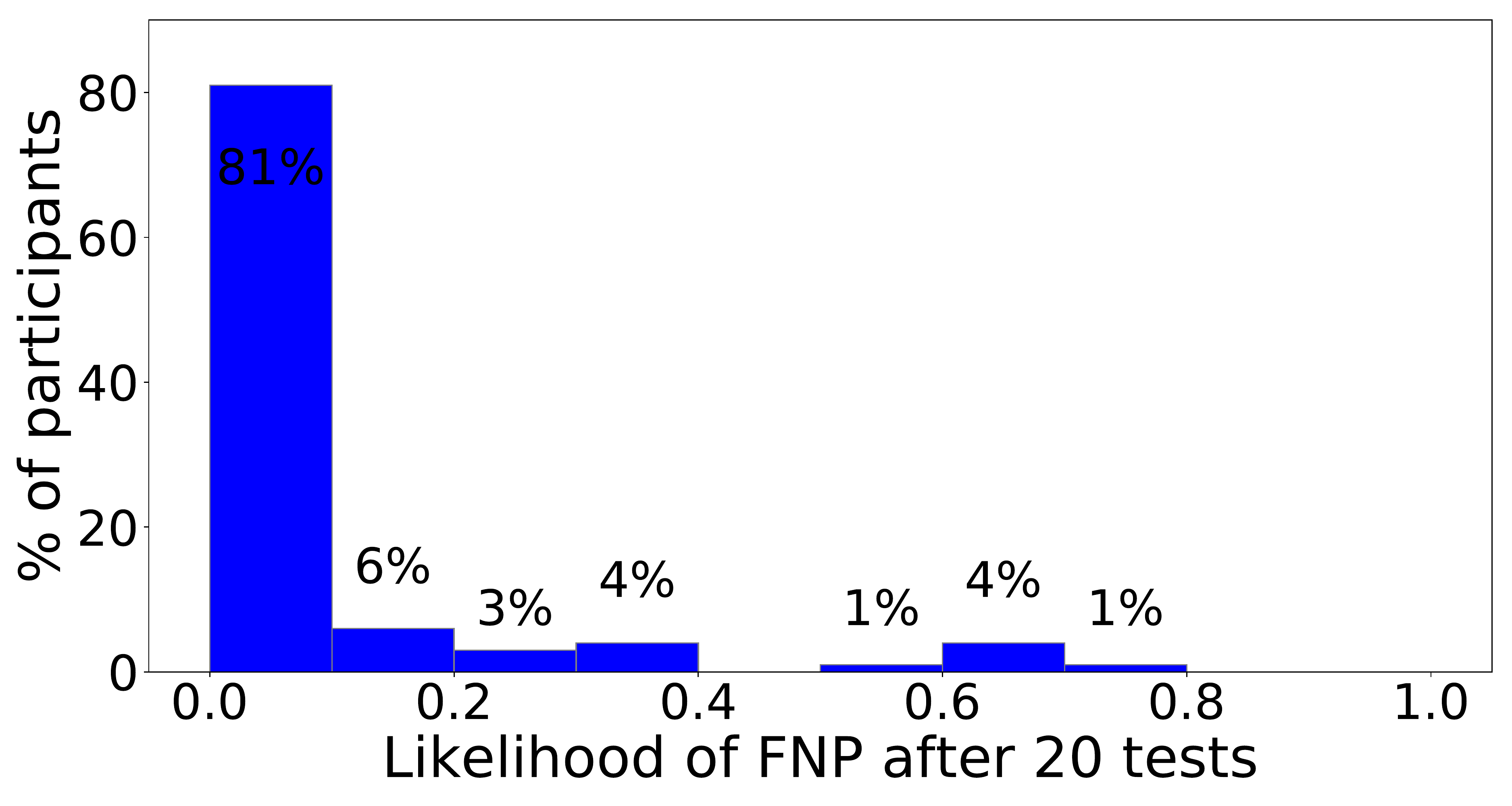}
	\end{subfigure}		
	\caption{The cancer risk prediction scenario---the number of participants matched with each notion of fairness (y-axis) along with the likelihood levels (x-axis). Demographic parity captures the choices made by the majority of participants. }\label{fig:histogram_med}
\end{figure*}

\begin{figure*}[t!]
	\centering
	\begin{subfigure}[b]{0.48\textwidth}
		\includegraphics[width=\textwidth]{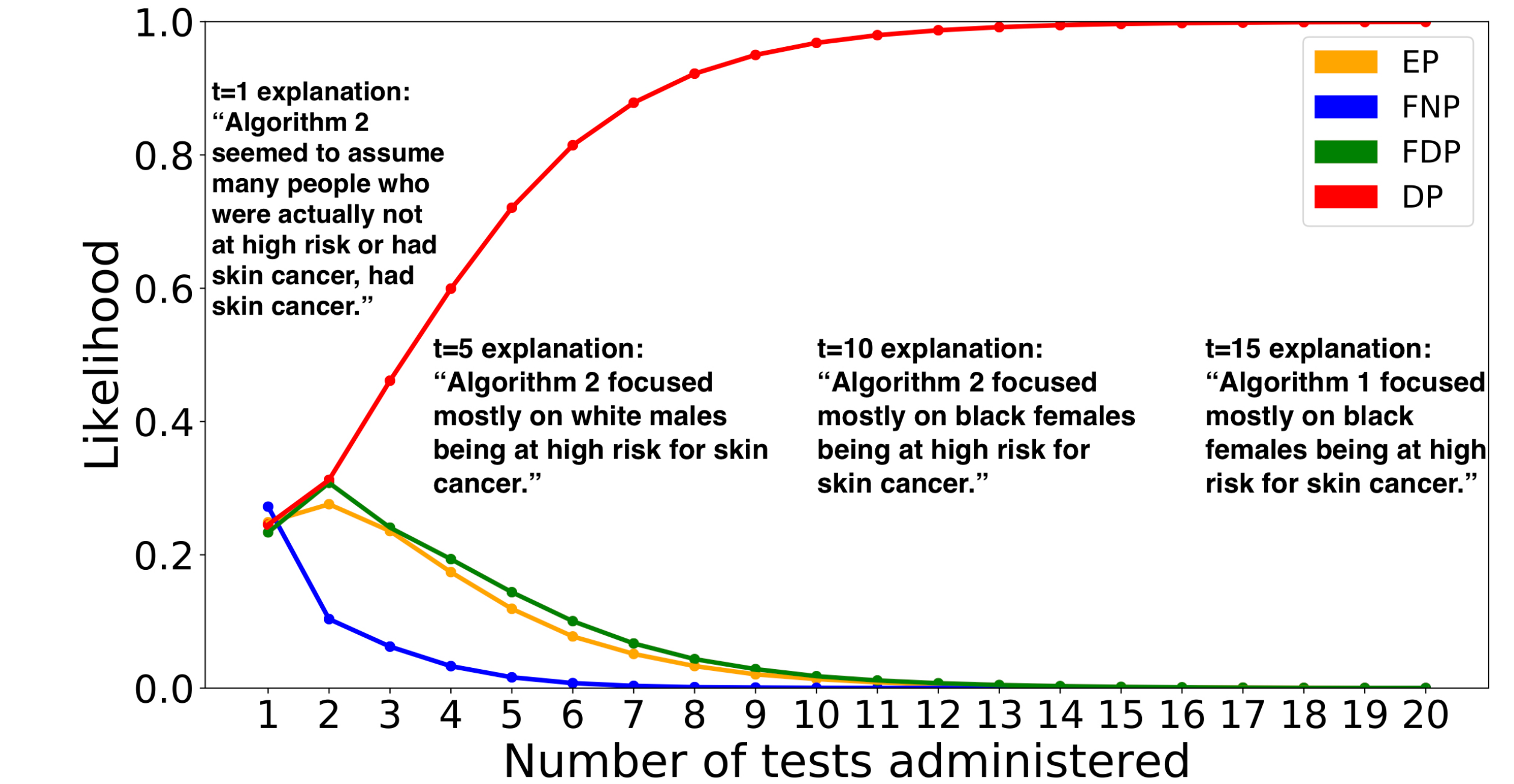}
	\end{subfigure}
	\begin{subfigure}[b]{0.48\textwidth}
		\includegraphics[width=\textwidth]{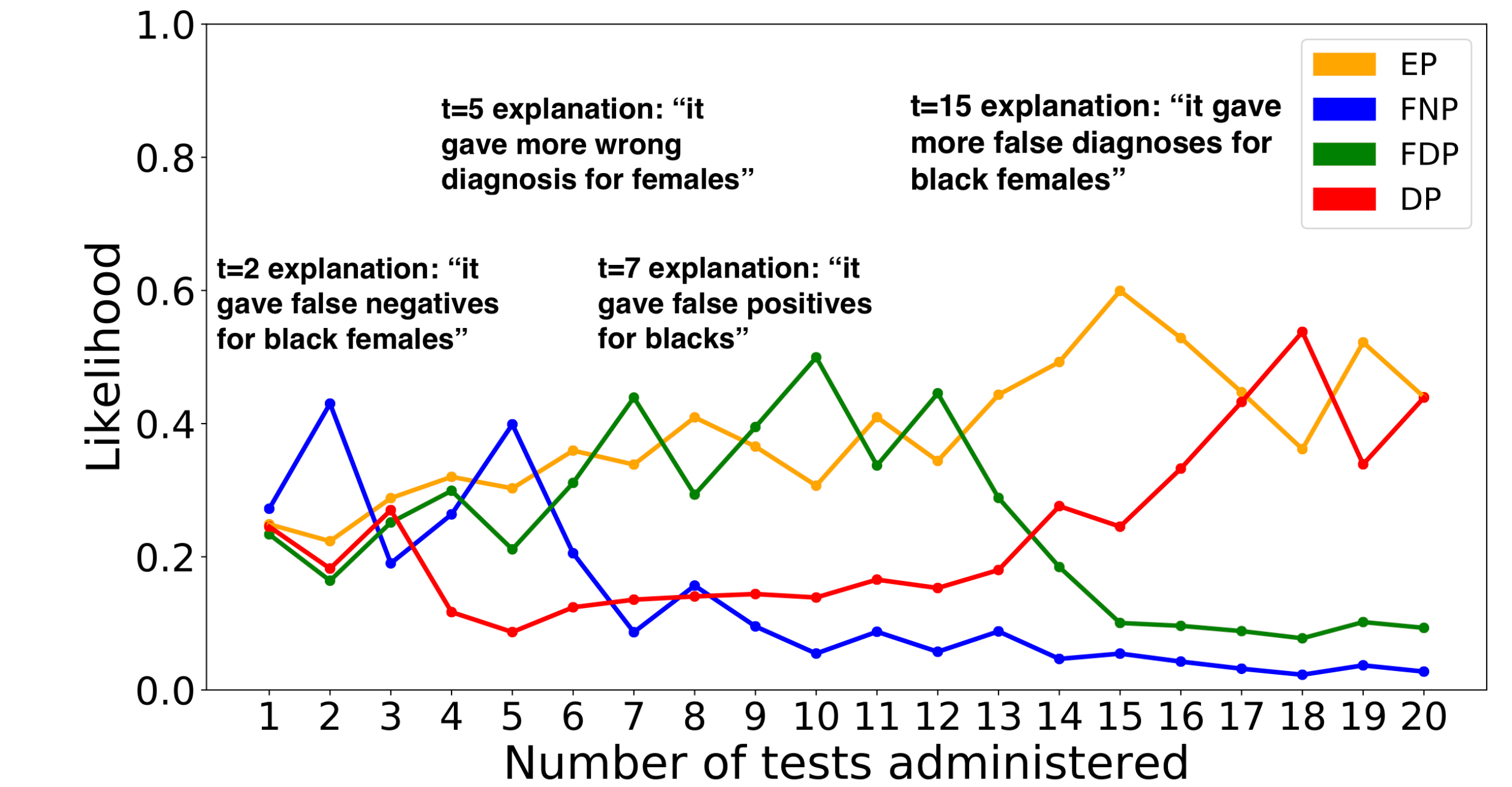}
	\end{subfigure}
	\caption{Trajectories of two different participants, one with a high final likelihood on DP and one with a low final likelihood on any fairness notion. Explanations provided by the former participant demonstrates consistency with an explanation aligned with demographic parity---addressing the predictions without considering ground truth labels. Explanations provided by the latter participant demonstrates inconsistency across different fairness notions.}\label{fig:trajectory}
\end{figure*}

%\begin{figure}[t!]
%	\centering
%	\begin{subfigure}[b]{0.24\textwidth}
%		\includegraphics[width=\textwidth]{Figures/Dropdown_Pilot_Crime/pilot_crime_dp.pdf}
%	\end{subfigure}
%	\begin{subfigure}[b]{0.24\textwidth}
%		\includegraphics[width=\textwidth]{Figures/Dropdown_Pilot_Med/pilot_med_dp.pdf}
%	\end{subfigure}	
%	\caption{Histogram specifying the number of participants matched with demographic parity when participants are required to pick a structured explanation.}\label{fig:drop_down}
%\end{figure}

\subsection{Qualitative Results}
Table~\ref{tab:explanation} shows several instances of the explanations provided by the participants in each category. For example, a participant categorized by our system as following the Error Parity hypothesis provided an explanation that focused on  the accuracy rate within demographic groups. These explanations demonstrate the ability for human participants to provide explanations in free text that can align with mathematical notions of fairness, without being prompted to think along these statistical definitions. 

\begin{table}[t!]
\caption{Typical explanations provided by participants in each category.}
\label{tab:explanation}
\centering
\begin{tabularx}{0.48\textwidth}{l  X }
Category & Explanation\\ \hline \hline
DP  &  \textit{``It expected only black females to reoffend.''} \\ 
EP  &  \textit{``Algorithm 1 made more correct decisions for white people.''} \\ 
FDR  &  \textit{``White males are incorrectly labeled likely to reoffend more often by this algorithm than the other.''} \\ 
FNR  &  \textit{``Two white males who did reoffend, both received \emph{will not reoffend} status.'' } \\ \hline
\end{tabularx}
\end{table}

The free text explanations provided by participants present us with insight into their trajectories, and why a small number of participants could not be confidently categorized as either one of the four notions. Figure~\ref{fig:trajectory} compares the trajectories of two participants - one categorized as following Demographic Parity with high confidence, and another participant who is not confidently categorized into any metric. The latter participant's explanations vary across time steps, from considering false negative rates to accuracy within demographic groups. These inconsistent explanations support our system's inability to assign a fairness notion with high confidence. 

%Do participants pick their notion of fairness during the first few questions, and thereafter follow that notion? 

%\begin{table*}[t!]
%\caption{Instances of explanations provided by participants who were not confidently categorized.}
%\label{tab:none}
%\begin{tabular}{l l l l}
%Instance & Explanation\\ \hline \hline
%1  &  \hh{placeholder} \\ 
%2  &  \hh{placeholder} \\ 
%3  &  \hh{placeholder} \\ 
%4  &  \hh{placeholder} \\ 
%5 &  \hh{placeholder} \\ \hline
%\end{tabular}
%\end{table*}

%Most of the scenarios seem pretty similar. very difficult to discern. It seemed very repetitive.
%
%sliding scale option 
%there could be a "neither" option
%
%Hard to label things discriminatory when all I know is race and gender. I hope the AI was making decisions on more than that.

%I would have liked to know age and education to see if those were also factors though.
%I wish I would've had a little more variety in algorithm choices or more information about how the information was accessed when making decision.
% The question must come up: is it discriminatory to predict correctly? This is something to mull over.

\section{Survey Design and Analysis} \label{sec:survey}

To test \textbf{H3}, we presented participants with three algorithms, each offering a different trade-off between accuracy and equality. We asked our participants to choose the one they consider ethically more desirable. 
For the case of medical risk prediction, we performed this survey for two different scenarios: 1) predicting the risk of skin cancer (high-stakes), and predicting the severity of flu symptoms (low stakes). We hypothesized that when the stakes are high, more participants would choose high overall accuracy ( i.e., $A_1$) over low inequality (i.e., $A_3$). Similarly, for the case of criminal risk assessment, we conducted the survey for both a high-stakes (predictions used to determine jail time) and a low-stakes scenario (predictions used to set bail amount.) Below we lay out the questionnaires precisely as they were presented to the survey participants.

\begin{table}[h]
\caption{Three hypothetical algorithms offering distinct tradeoffs between accuracy and inequality.}
\label{tab:acc}
\centering
\begin{tabular}{lllll}
Algorithm &  accuracy &  female acc. &  male acc. \\ \hline
$A_1$ & 94\% & 89\% &  99\% \\
$A_2$ & 91\% & 90\% &  92\% \\
$A_3$ & 86\% & 86\% &  86\% 
\end{tabular}
\end{table}

\paragraph{\textbf{Skin cancer risk prediction}} \textit{Data-driven algorithms are increasingly employed to screen and predict the risk of various forms of diseases, such as skin cancer. They can find patterns and links in medical records that previously required great levels of expertise and time from human doctors. Algorithmic predictions are then utilized by health-care professionals to create the appropriate treatment plans for patients.
Suppose we have three skin cancer risk prediction algorithms and would like to decide which one should be deployed for cancer screening of patients in a hospital. Each algorithm has a specific level of accuracy---where accuracy specifies the percentage of subjects for whom the algorithm makes a correct prediction. See Table~\ref{tab:acc}. 
Note that in cases where the deployed algorithm makes an error, a patient's life can be severely impacted. A patient falsely predicted to have high risk of skin cancer may unnecessarily undergo high-risk and costly medical interventions, while a patient falsely labeled as low risk for skin cancer may face a significantly lower life expectancy.
From a moral standpoint, which one of the following three algorithms do you think is more desirable for deployment in real-world hospitals? }

\paragraph{\textbf{Flu symptom severity prediction}} \textit{Data-driven algorithms can be employed to screen and predict the risk of various forms of diseases, including common flu. They can find patterns and links in medical records that previously required visiting a human doctor. Algorithmic predictions can be utilized by patients to decide whether to see a doctor for their symptoms.
Suppose we have three different algorithms predicting the severity of flu symptoms in patients and would like to decide which one should be deployed in the real world. Each algorithm has a specific level of accuracy---where accuracy specifies the percentage of subjects for whom the algorithm makes a correct prediction. 
Note that in cases where the deployed algorithm makes an error, a patient will be temporarily impacted negatively. A patient falsely predicted to develop severe flu symptoms may unnecessarily seek medical intervention, while a patient falsely labeled as developing only mild symptoms may have to cope with severe symptoms for a longer period of time (at most two weeks).
}

\begin{figure}[h!]
	\centering
	\includegraphics[width=0.48\textwidth]{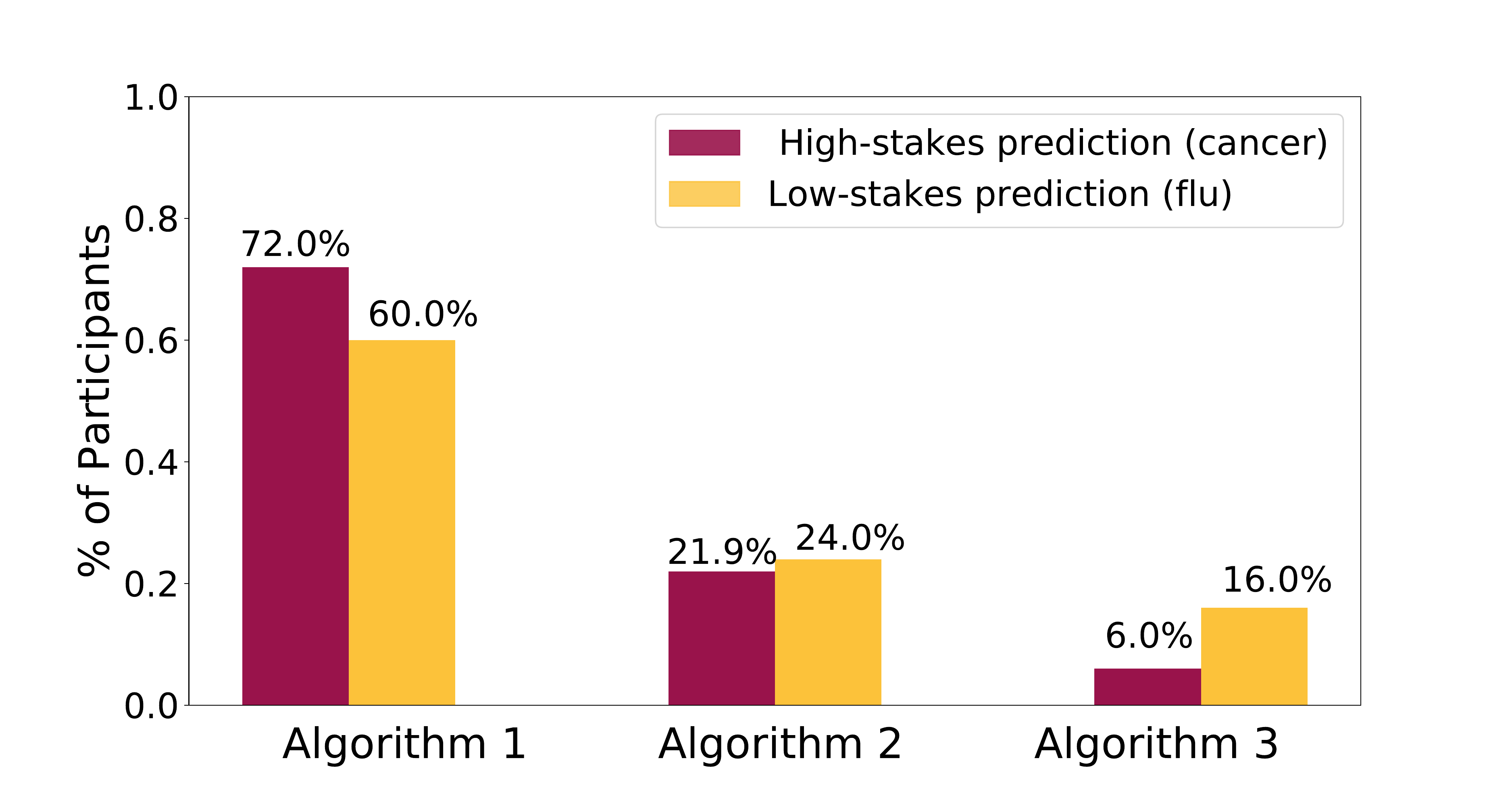}
	\caption{Medical risk prediction scenarios. Participants gave higher weight to accuracy (compared to inequality) when predictions can impact patients' life expectancy.}
	\label{fig:survey_med}
\end{figure}

\begin{figure}[h!]
	\centering
	\includegraphics[width=0.48\textwidth]{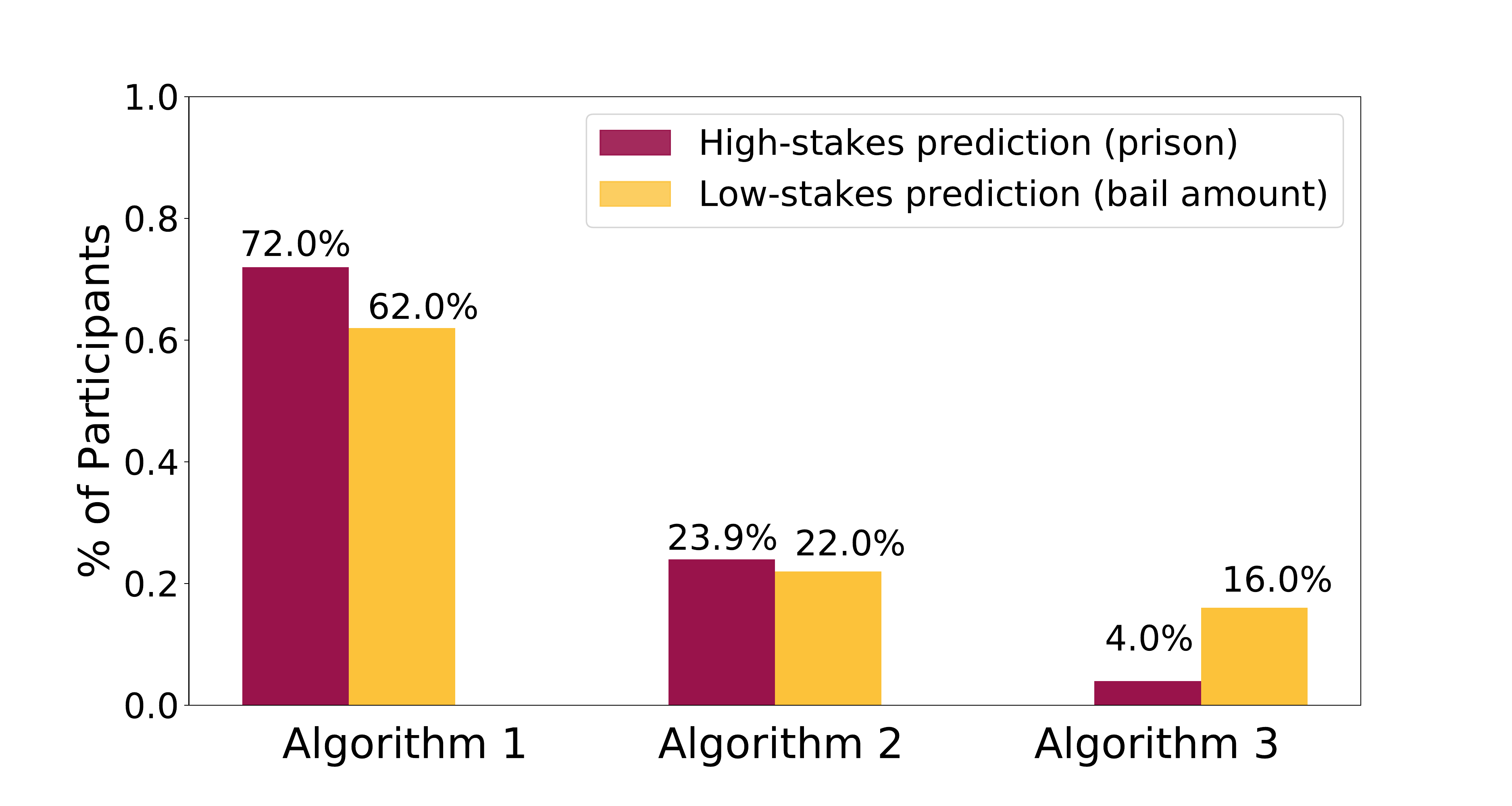}
	\caption{Crime risk prediction scenarios. Participants gave higher weight to accuracy (compared to inequality) when predictions can impact defendants' life trajectory.}
	\label{fig:survey_crime}
\end{figure}

 \begin{table*}[t!]
\caption{Examples of participant feedback demonstrating positive educational effect of the task.}
\label{tab:feedback}
\centering
\begin{tabularx}{\textwidth}{l  X }
Instance & Feedback\\ \hline \hline
1  &  \textit{``It was more difficult than I first thought to choose an algorithm. The question must come up: is it discriminatory to predict correctly? This is something to mull over. Thank you!''} \\ 
2  &  \textit{``I would have liked to know age and education to see if those were also factors though.''} \\ 
3  &  \textit{``This was really fun, thank you! I've never realized that algorithms were used for this sort of data, I hope this helps in some way to improve those systems.''} \\ 
4  &  \textit{``The justice system's algorithms think very poorly of white men and black women.  That's frightening if true.''} \\ 
5 &  \textit{``I thought this was really interesting! I'd be curious what the results say.''} \\ \hline
\end{tabularx}
\end{table*}

Figures~\ref{fig:survey_crime} and \ref{fig:survey_med} show the number of participants who chose each algorithm. As hypothesized, when the stakes are high, participants gave higher weight to accuracy and lower weight to inequality.

%\vspace{-3mm}

\section{Discussion}\label{sec:discussion}

%\subsection{Lessons and Insights}
The main takeaway message from our experiments is that \textbf{demographic parity best captures people's perception of fairness.} Our surveys show that \textbf{participants consider accuracy more important than equality when stakes are high.}
Finally, participants on Amazon Mechanical Turk found the task engaging and informative, while some thought more context about the algorithm and decision subjects would have helped them make a more informed choice. Feedback from the users, such as the examples shown in Table~\ref{tab:feedback}, demonstrate that our task encouraged users to think about what factors algorithmic decision-making systems should take into account, as well as reflect about the use of discriminatory algorithms in society.

\subsection{Limitations}
The primary goal of our work was to find out which existing notion of fairness best captures lay people's perception of fairness in a given context. We took it as a given that at least one of these notions is a good representation of human judgment---at least in the highly stylized setting we presented them with. We acknowledge that real-world scenarios are always much more complex and there are many factors (beyond true and predicted labels) that impact people's judgment of the situation. Existing mathematical notions of fairness are in comparison very simplistic, and they can never reflect all the nuances involved. Our work is by no means a final verdict---it is an initial step toward a better understanding of fairness and justice from the perspective of everyday people.

%We assumed people focus solely on inequality.

One barrier to obtaining meaningful answers from participants is engagement. On AMT it is difficult to monitor participants' attention to the task.
We were particularly concerned about the possibility of participants choosing their answers without careful consideration of all factors in front of them---and in the worst case, completely at random. We prevented this by 
\begin{itemize}
\item Restricting participation to turkers with $99\%$ approval rate.
\item Limiting the number of tests to at most 20.
\item Requiring the participant to provide an explanation for their choice.
\item Restricting participants to complete the task at most once.
\end{itemize}

As with any experiment, we cannot completely rule out the potential impact of framing. We experimented with two different UIs, and our findings were robust to that choice. This, however, does not mean that a different experimental setting could not lead to different conclusions. In particular, in our experiments, participants did not have personal stakes in their choice of algorithm, and they might have responded differently if they could be the subject of decision making through their choice of algorithm. 

\hh{In all of our tests, we restricted attention to two predictive models of similar accuracy. How do participants' perception of discrimination change if the two models presented to them differ in terms of accuracy? We leave the study of the role of accuracy for future work.}
%limited attention to tests where the two algorithms exhibit equal overall accuracy

\subsection{Future Directions}
Our findings have important implications for the Fair ML literature. In particular, they accentuate the need for a more comprehensive understanding of the human attitude toward algorithmic fairness. Algorithmic decisions will ultimately impact human subjects' lives, and it is, therefore, critical to involve them in the process of choosing the right notion of fairness. Our work takes an initial step toward this broader agenda. Directions for future work include, but are not limited to, the following:
\begin{itemize}
\item providing subjects with more information about subjects of algorithmic decision making and inner workings of each algorithm; quantifying how this additional context affects their assessment of algorithmic fairness.
\item providing subjects with information about the non-algorithmic alternatives (i.e., cost, accuracy, and bias of human decisions)
\item exploring fairness considerations at different levels; are people more sensitive to individual- or group-level unfairness?
\item studying the role of personal stakes; do people assess algorithmic fairness differently if they may be personally affected by algorithmic decisions? 
\item \hh{studying the effect of participant's expertise; does expertise in law and discrimination qualitatively change the participants' responses?}
\end{itemize}

\iffalse
b) Conducting unbiased surveys is very difficult, if not impossible (bias can occur during selection of users, the questions asked and how they are phrased, non-response, etc.). Also, it's worth reflecting on whether the majority opinion is desirable in all cases (e.g., in the case of decisions affecting marginalized and small minorities). Further, active learning has the potential to inflate the bias due to propagation of positive correlations.

I would have loved to see this study to also include individual fairness or calibration.

The scaling protocol seems good, but it would be really nice to see this extended to a much larger population size to verify the results. Another very interesting study would be to compare this with a study on a set of legal experts. 
\fi

\section{Acknowledgments}
%H. Heidari and A. Krause acknowledge support from the CTI grant no. 27248.1 PFES-ES. 
M. Srivastava acknowledges support from the National Science Foundation Graduate Research Fellowship Program under Grant No. DGE-1656518 and from the ETH Z{\"u}rich Student Summer Research Fellowship. Any opinions, findings, and conclusions or recommendations expressed in this material are those of the author(s) and do not necessarily reflect the views of the funding agencies. 

\bibliographystyle{named}
\bibliography{MyBib}

\begin{thebibliography}{}

\bibitem[\protect\citeauthoryear{Angwin \bgroup \em et al.\egroup
  }{2016}]{propublica}
Julia Angwin, Jeff Larson, Surya Mattu, and Lauren Kirchner.
\newblock Machine bias.
\newblock {\em Propublica}, 2016.

\bibitem[\protect\citeauthoryear{Awad \bgroup \em et al.\egroup
  }{2018}]{awad2018moral}
Edmond Awad, Sohan Dsouza, Richard Kim, Jonathan Schulz, Joseph Henrich, Azim
  Shariff, Jean-Fran{\c{c}}ois Bonnefon, and Iyad Rahwan.
\newblock The moral machine experiment.
\newblock {\em Nature}, 563(7729):59, 2018.

\bibitem[\protect\citeauthoryear{Barry-Jester \bgroup \em et al.\egroup
  }{2015}]{sentencing}
Anna Barry-Jester, Ben Casselman, and Dana Goldstein.
\newblock The new science of sentencing.
\newblock {\em The Marshall Project}, August 2015.

\bibitem[\protect\citeauthoryear{Binns \bgroup \em et al.\egroup
  }{2018}]{binns2018s}
Reuben Binns, Max Van~Kleek, Michael Veale, Ulrik Lyngs, Jun Zhao, and Nigel
  Shadbolt.
\newblock 'it's reducing a human being to a percentage': Perceptions of justice
  in algorithmic decisions.
\newblock In {\em CHI}, page 377. ACM, 2018.

\bibitem[\protect\citeauthoryear{Buolamwini and
  Gebru}{2018}]{buolamwini2018gender}
Joy Buolamwini and Timnit Gebru.
\newblock Gender shades: Intersectional accuracy disparities in commercial
  gender classification.
\newblock In {\em In Proceedings of the 1st Conference on Fairness,
  Accountability and Transparency (FAT*)}, pages 77--91, 2018.

\bibitem[\protect\citeauthoryear{Chouldechova}{2017}]{chouldechova2017fair}
Alexandra Chouldechova.
\newblock Fair prediction with disparate impact: A study of bias in recidivism
  prediction instruments.
\newblock {\em arXiv preprint arXiv:1703.00056}, 2017.

\bibitem[\protect\citeauthoryear{Corbett-Davies \bgroup \em et al.\egroup
  }{2017}]{corbett2017algorithmic}
Sam Corbett-Davies, Emma Pierson, Avi Feller, Sharad Goel, and Aziz Huq.
\newblock Algorithmic decision making and the cost of fairness.
\newblock In {\em KDD}, pages 797--806. ACM, 2017.

\bibitem[\protect\citeauthoryear{Deo}{2015}]{deo2015machine}
Rahul~C. Deo.
\newblock Machine learning in medicine.
\newblock {\em Circulation}, 132(20):1920--1930, 2015.

\bibitem[\protect\citeauthoryear{Dwork \bgroup \em et al.\egroup
  }{2012}]{dwork2012fairness}
Cynthia Dwork, Moritz Hardt, Toniann Pitassi, Omer Reingold, and Richard Zemel.
\newblock Fairness through awareness.
\newblock In {\em Proceedings of the Innovations in Theoretical Computer
  Science Conference (ITCS)}, pages 214--226. ACM, 2012.

\bibitem[\protect\citeauthoryear{Feldman \bgroup \em et al.\egroup
  }{2015}]{feldman2015certifying}
Michael Feldman, Sorelle~A. Friedler, John Moeller, Carlos Scheidegger, and
  Suresh Venkatasubramanian.
\newblock Certifying and removing disparate impact.
\newblock In {\em KDD}, pages 259--268. ACM, 2015.

\bibitem[\protect\citeauthoryear{Gajane and
  Pechenizkiy}{2017}]{gajane2017formalizing}
Pratik Gajane and Mykola Pechenizkiy.
\newblock On formalizing fairness in prediction with machine learning.
\newblock {\em arXiv preprint arXiv:1710.03184}, 2017.

\bibitem[\protect\citeauthoryear{Golovin \bgroup \em et al.\egroup
  }{2010}]{golovin2010near}
Daniel Golovin, Andreas Krause, and Debajyoti Ray.
\newblock Near-optimal bayesian active learning with noisy observations.
\newblock In {\em NIPS}, pages 766--774, 2010.

\bibitem[\protect\citeauthoryear{Grgic-Hlaca \bgroup \em et al.\egroup
  }{2018}]{grgic2018human}
Nina Grgic-Hlaca, Elissa~M. Redmiles, Krishna~P Gummadi, and Adrian Weller.
\newblock Human perceptions of fairness in algorithmic decision making: A case
  study of criminal risk prediction.
\newblock In {\em WWW}, pages 903--912, 2018.

\bibitem[\protect\citeauthoryear{Hardt \bgroup \em et al.\egroup
  }{2016}]{hardt2016equality}
Moritz Hardt, Eric Price, and Nati Srebro.
\newblock Equality of opportunity in supervised learning.
\newblock In {\em NIPS}, pages 3315--3323, 2016.

\bibitem[\protect\citeauthoryear{Hart}{2017}]{medicine}
Robert~David Hart.
\newblock If you're not a white male, artificial intelligence's use in
  healthcare could be dangerous.
\newblock {\em Quartz}, July 2017.

\bibitem[\protect\citeauthoryear{Heidari \bgroup \em et al.\egroup
  }{2019}]{heidari2019a}
Hoda Heidari, Michele Loi, Krishna~P. Gummadi, and Andreas Krause.
\newblock A moral framework for understanding of fair ml through economic
  models of equality of opportunity.
\newblock In {\em FAT*}, 2019.

\bibitem[\protect\citeauthoryear{Holstein \bgroup \em et al.\egroup
  }{2018}]{holstein2018improving}
Kenneth Holstein, Jennifer~Wortman Vaughan, Hal Daum{\'e}~III, Miro Dud{\'\i}k,
  and Hanna Wallach.
\newblock Improving fairness in machine learning systems: What do industry
  practitioners need?
\newblock In {\em CHI}, 2018.

\bibitem[\protect\citeauthoryear{Joseph \bgroup \em et al.\egroup
  }{2016}]{joseph2016fairness}
Matthew Joseph, Michael Kearns, Jamie Morgenstern, and Aaron Roth.
\newblock Fairness in learning: Classic and contextual bandits.
\newblock In {\em NIPS}, pages 325--333, 2016.

\bibitem[\protect\citeauthoryear{Kleinberg \bgroup \em et al.\egroup
  }{2017}]{kleinberg2016inherent}
Jon Kleinberg, Sendhil Mullainathan, and Manish Raghavan.
\newblock Inherent trade-offs in the fair determination of risk scores.
\newblock In {\em ITCS}, 2017.

\bibitem[\protect\citeauthoryear{Lee and Baykal}{2017}]{lee2017algorithmic}
Min~Kyung Lee and Su~Baykal.
\newblock Algorithmic mediation in group decisions: Fairness perceptions of
  algorithmically mediated vs. discussion-based social division.
\newblock In {\em CSCW}, pages 1035--1048, 2017.

\bibitem[\protect\citeauthoryear{Lee \bgroup \em et al.\egroup
  }{2018}]{lee2018webuildai}
Min~Kyung Lee, Daniel Kusbit, Anson Kahng, Ji~Tae Kim, Xinran Yuan, Allissa
  Chan, Ritesh Noothigattu, Daniel See, Siheon Lee, and Christos-Alexandros
  Psomas.
\newblock Webuildai: Participatory framework for fair and efficient algorithmic
  governance, 2018.

\bibitem[\protect\citeauthoryear{Noothigattu \bgroup \em et al.\egroup
  }{2018}]{noothigattu2018voting}
Ritesh Noothigattu, Snehalkumar 'Neil'~S. Gaikwad, Edmond Awad, Sohan Dsouza,
  Iyad Rahwan, Pradeep Ravikumar, and Ariel~D. Procaccia.
\newblock A voting-based system for ethical decision making.
\newblock In {\em AAAI}, 2018.

\bibitem[\protect\citeauthoryear{Petrasic \bgroup \em et al.\egroup
  }{2017}]{whitecase}
Kevin Petrasic, Benjamin Saul, James Greig, and Matthew Bornfreund.
\newblock Algorithms and bias: What lenders need to know.
\newblock {\em White \& Case}, 2017.

\bibitem[\protect\citeauthoryear{Ray \bgroup \em et al.\egroup
  }{2012}]{ray2012bayesian}
Debajyoti Ray, Daniel Golovin, Andreas Krause, and Colin Camerer.
\newblock Bayesian rapid optimal adaptive design (broad): Method and
  application distinguishing models of risky choice.
\newblock {\em California Institute of Technology working paper}, 2012.

\bibitem[\protect\citeauthoryear{Rudin}{2013}]{policing}
Cynthia Rudin.
\newblock Predictive policing using machine learning to detect patterns of
  crime.
\newblock {\em Wired Magazine}, August 2013.
\newblock Retrieved 4/28/2016.

\bibitem[\protect\citeauthoryear{Saxena \bgroup \em et al.\egroup
  }{2018}]{saxena2018fairness}
Nripsuta Saxena, Karen Huang, Evan DeFilippis, Goran Radanovic, David Parkes,
  and Yang Liu.
\newblock How do fairness definitions fare? examining public attitudes towards
  algorithmic definitions of fairness.
\newblock {\em arXiv preprint arXiv:1811.03654}, 2018.

\bibitem[\protect\citeauthoryear{Speicher \bgroup \em et al.\egroup
  }{2018}]{speicher2018a}
Till Speicher, Hoda Heidari, Nina Grgic-Hlaca, Krishna~P. Gummadi, Adish
  Singla, Adrian Weller, and Muhammad~Bilal Zafar.
\newblock A unified approach to quantifying algorithmic unfairness: Measuring
  individual and group unfairness via inequality indices.
\newblock In {\em KDD}, 2018.

\bibitem[\protect\citeauthoryear{Sweeney}{2013}]{sweeney2013discrimination}
Latanya Sweeney.
\newblock Discrimination in online ad delivery.
\newblock {\em Queue}, 11(3):10, 2013.

\bibitem[\protect\citeauthoryear{Taylor and
  Wright}{2005}]{taylor2005importance}
Anne Taylor and Jackson Wright.
\newblock Importance of race/ethnicity in clinical trials.
\newblock {\em Circulation}, 112(23):3654--3660, 2005.

\bibitem[\protect\citeauthoryear{Veale \bgroup \em et al.\egroup
  }{2018}]{veale2018fairness}
Michael Veale, Max Van~Kleek, and Reuben Binns.
\newblock Fairness and accountability design needs for algorithmic support in
  high-stakes public sector decision-making.
\newblock In {\em CHI}, page 440. ACM, 2018.

\bibitem[\protect\citeauthoryear{Woodruff \bgroup \em et al.\egroup
  }{2018}]{woodruff2018qualitative}
Allison Woodruff, Sarah~E. Fox, Steven Rousso-Schindler, and Jeffrey Warshaw.
\newblock A qualitative exploration of perceptions of algorithmic fairness.
\newblock In {\em CHI}, page 656. ACM, 2018.

\bibitem[\protect\citeauthoryear{Zafar \bgroup \em et al.\egroup
  }{2017}]{zafar2017dmt}
Muhammad~Bilal Zafar, Isabel Valera, Manuel Gomez~Rodriguez, and Krishna~P.
  Gummadi.
\newblock Fairness beyond disparate treatment \& disparate impact: Learning
  classification without disparate mistreatment.
\newblock In {\em WWW}, pages 1171--1180, 2017.

\end{thebibliography}

% At most two pages, focus on reproducability.

\section{Reproducibility}

\subsection{Task Interface}
%The interface that we provided to Amazon Mechanical Turk participants is available at: \url{http://165.227.25.219:8080/?uiversion=v2&userid=example_uid}.
Each of the 20 tests were adaptively selected based on the participant's answers to previous tests, with the first test provided at random. A return code was provided for AMT participants to give back to us after they finished all 20 tests---this allowed us to link AMT participants survey responses (i.e., demographic information and feedback) with the logs of server interactions that we maintained. 

%While this interface is currently designed for the purpose of data collection, the interface we will release to the general public will be the same except for the final page: Instead of a return code, we intend to output the likelihoods across fairness notions, so that readers and the general public can use it as an educational tool.  

\subsection{Code}
The code for our server, implementation of adaptive test selection, and analysis can all be found on our Git Repository: \url{https://github.com/meghabyte/FairnessPerceptions/}. A more detailed description of how to run the code and reproduce the plots in our paper can be found in the repository's ReadMe.  A very brief description follows: 

\paragraph{\textbf{Amazon Mechanical Turk data}} All raw data from our Amazon Mechanical Turk experiments can be found under the data folder in the repository. These consist of server logs tracking all interactions with the server. We also provide the following processed data files:
\begin{itemize}
\item A file combining each participant's data with their likelihoods across hypotheses/explanation after 20 tests;
\item  A file containing results from our pilot experiments (including the drop-down user interface that we chose to not use in our full experiments).
\end{itemize}
To respect the participants privacy, we do not release any data consisting of their identifying information, demographics, and feedback. 
\paragraph{\textbf{Analysis}} All plots from our paper, as well as the code to re-generate them, can be found in the analysis folder in the repository. 
\paragraph{\textbf{Interface}} Our interface can be run locally as well as a test mode where either random test selection (see Figure~\ref{fig:ec2_benefit}) or random answer generation (see Figure~\ref{fig:baseline}) can be simulated. All command line arguments are explained in the repository. 
%\\vfill\eject
\section{Additional Surveys}\label{app:survey}
\paragraph{\textbf{Sentencing time in prison}} \textit{Data-driven decision-making algorithms can be employed to predict the likelihood of future crime by defendants. These algorithmic predictions are utilized by judges to make sentencing decisions for defendants, in particular, how much time the defendant has to spend in prison.
Suppose we have three different algorithms predicting the risk of future crime for defendants and would like to decide which one should be deployed in real-world courtrooms. Each algorithm has a specific level of accuracy---where accuracy specifies the percentage of subjects for whom the algorithm makes a correct prediction.  See Table~\ref{tab:acc}. 
Note that in cases where the deployed algorithm makes an error, a defendant's life can be significantly impacted: A defendant falsely predicted to re-offend can unjustly face long prison sentences (minimum 1 year), while a defendant falsely predicted to not reoffend will not spend any time in prison and may commit a serious crime that could have been prevented.
From a moral standpoint, which one of the three algorithms do you think is more desirable for deployment in real-world courtrooms? 
}

\paragraph{\textbf{Setting the bail amount}} \textit{Data-driven decision-making algorithms can be employed to predict the likelihood of a defendant appearing in court for his/her future hearings. These algorithmic predictions can be utilized by judges to set the appropriate bail amount for the defendant.
Suppose we have three different algorithms predicting the risk of a defendant not showing up to future court hearings and we would like to decide which one should be deployed in real-world courtrooms. Each algorithm has a specific level of accuracy---where accuracy specifies the percentage of subjects for whom the algorithm makes a correct prediction. 
Note that in cases where the deployed algorithm makes an error, the defendant may be financially impacted: A defendant falsely predicted to not appear for future hearings is unnecessarily forced to submit a bail amount of ~\$2000. 
From a moral standpoint, which one of the three algorithms do you think is more desirable for deployment in real-world courtrooms? 
}

\section{Experiments with Other Group Fairness Notions}\label{app:additional}
For further experimentation, we replicated our study with the Demographic Parity (DP) excluded, and False Positive Rate Parity (FPP) and False Omission Rate Parity (FOP) included as hypotheses. We conducted this using the drop-down interface, and with 50 participants for the two settings (criminal recidivism prediction and skin cancer risk prediction). See Table~\ref{tab:extra_sum} for the summary. 

\begin{table}[h]
\caption{Number of participants matched with each notion with high likelihood ($>80\%$).}
\label{tab:extra_sum}
\centering
\begin{tabular}{lllllll}
  &  EP &  FPP &  FNP & FDR & FOP & None \\ \hline
Crime risk prediction &  0\% &  32\% &  18\% &  0\% &  0\%& 50\%\\
Cancer risk prediction & 28\% & 18\%  &  0\% & 2\%  &  0\%&  52\%\\
\end{tabular}
\end{table}

Without demographic parity included as a hypotheses, our system has a higher difficulty in matching a fairness hypotheses to a user with high likelihood - supporting our original experiment's findings that DP is the most common for users to choose. False Negatives and False Positives have the highest proportion of matches. More detailed plots of these additional experimental results are below in Figure~\ref{fig:histogram_crime_extra} and Figure~\ref{fig:histogram_med_extra}. 

\begin{figure*}[t!]
		\centering
		\begin{subfigure}[b]{0.31\textwidth}
			\includegraphics[width=\textwidth]{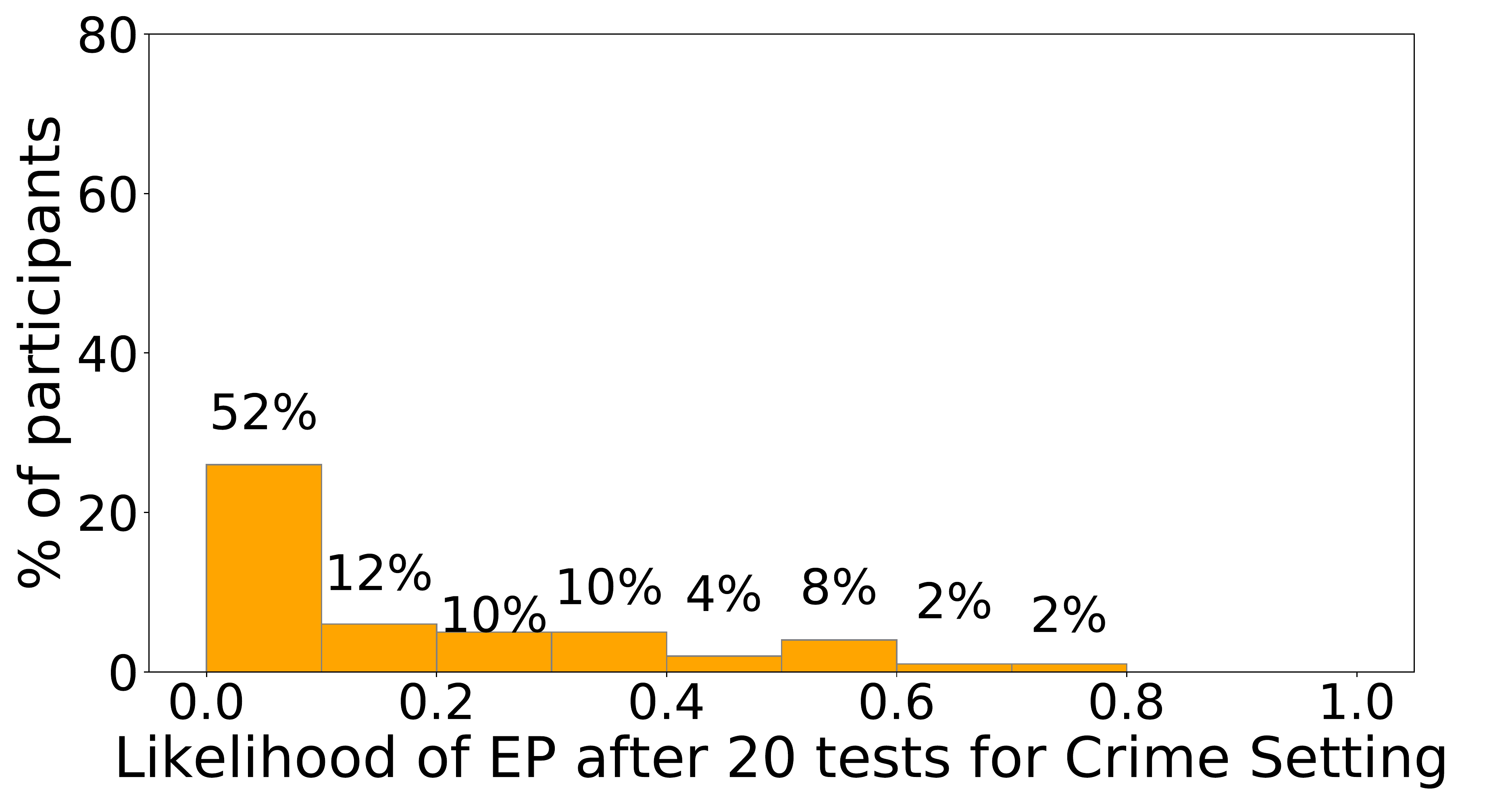}
		\end{subfigure}
		\begin{subfigure}[b]{0.31\textwidth}
			\includegraphics[width=\textwidth]{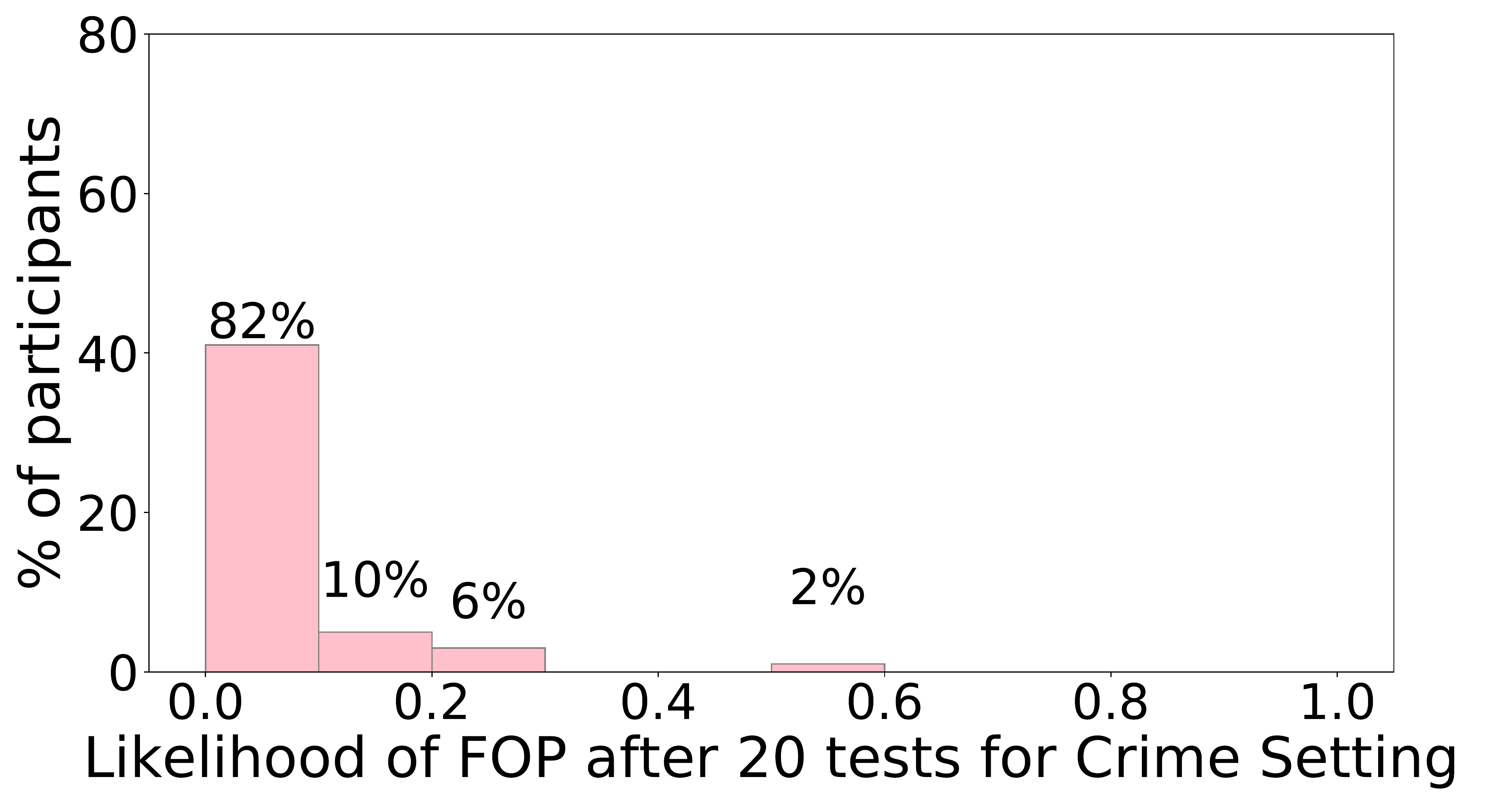}
		\end{subfigure}
		\begin{subfigure}[b]{0.31\textwidth}
			\includegraphics[width=\textwidth]{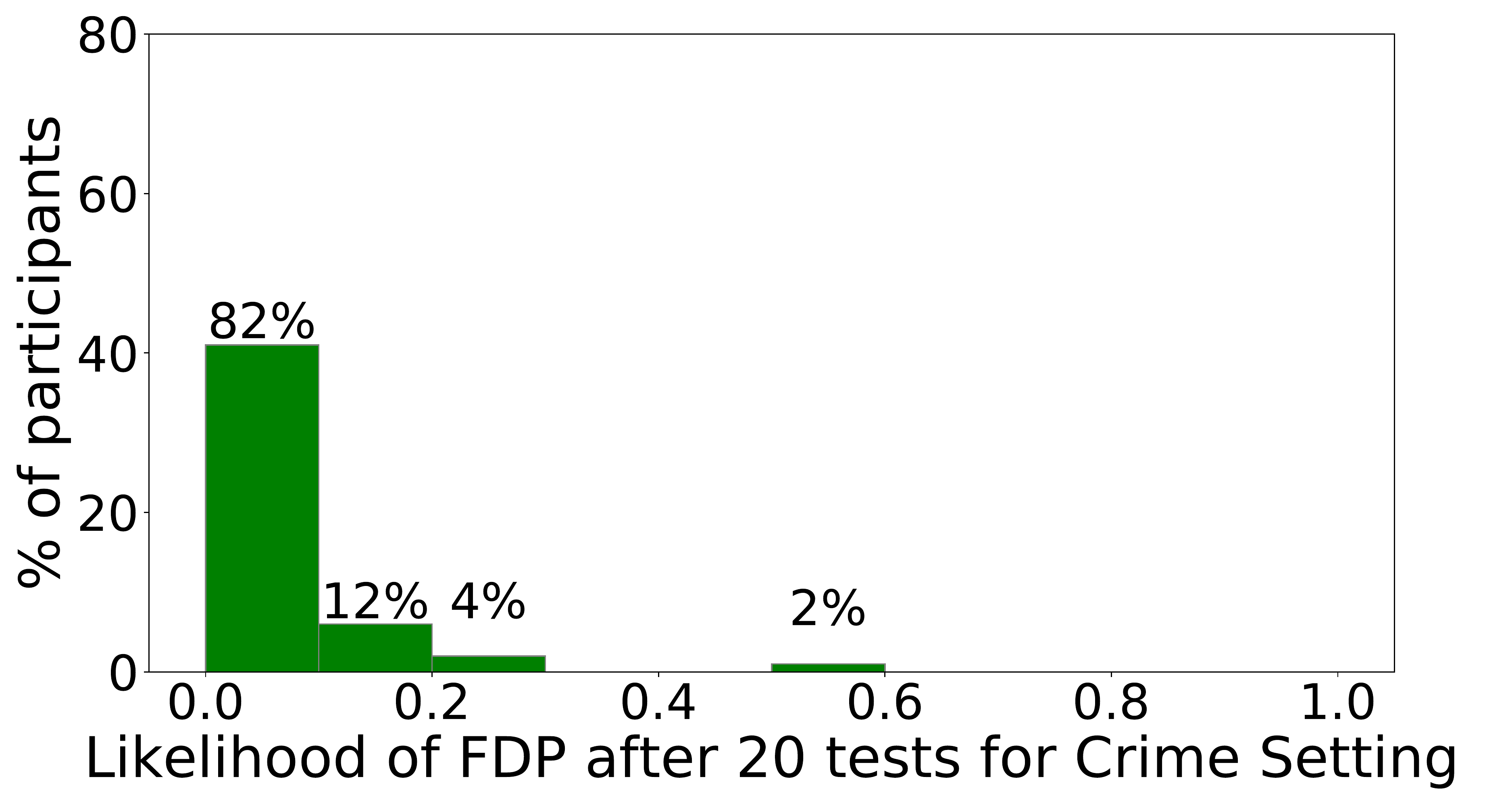}
		\end{subfigure}
		\\
		\begin{subfigure}[b]{0.31\textwidth}
			\includegraphics[width=\textwidth]{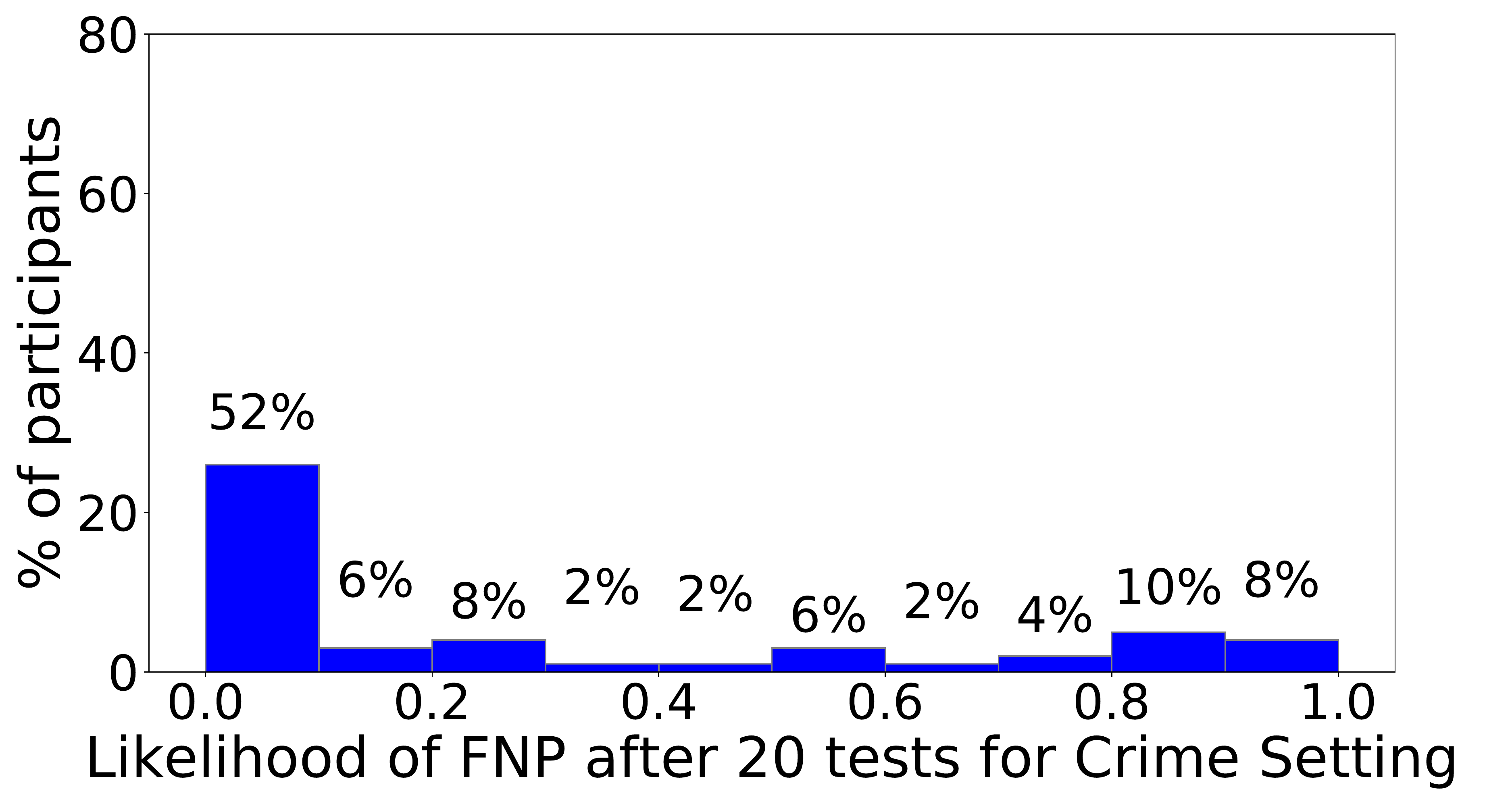}
		\end{subfigure}	  
		\begin{subfigure}[b]{0.31\textwidth}
			\includegraphics[width=\textwidth]{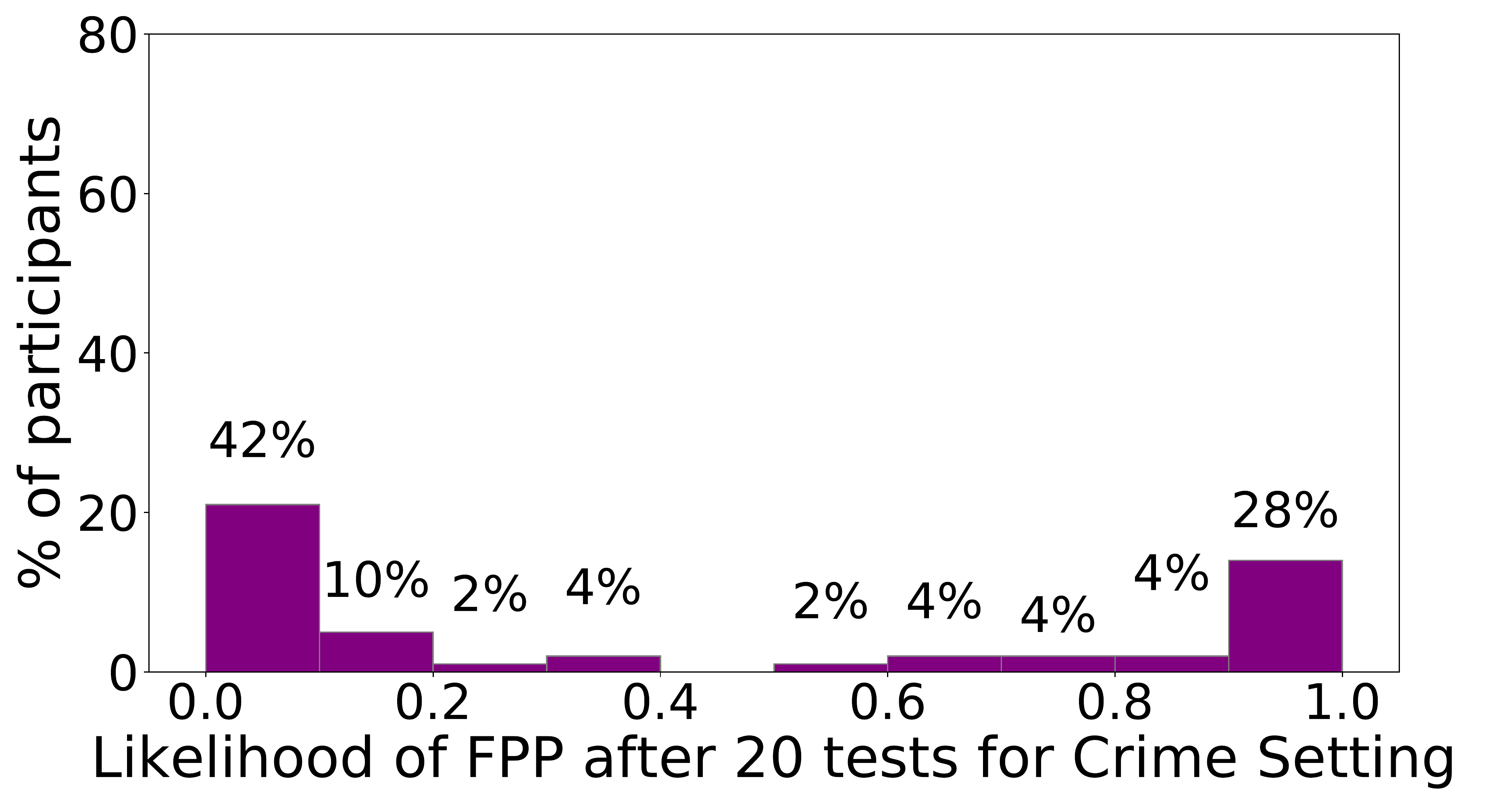}
		\end{subfigure}  
		\caption{The crime risk prediction scenario---the number of participants matched with each notion of fairness (y-axis) along with the likelihood levels (x-axis). False positive parity captures the choices made by the majority of participants. }\label{fig:histogram_crime_extra}
	\end{figure*} 
	
\begin{figure*}[t!]
		\centering
		\begin{subfigure}[b]{0.31\textwidth}
			\includegraphics[width=\textwidth]{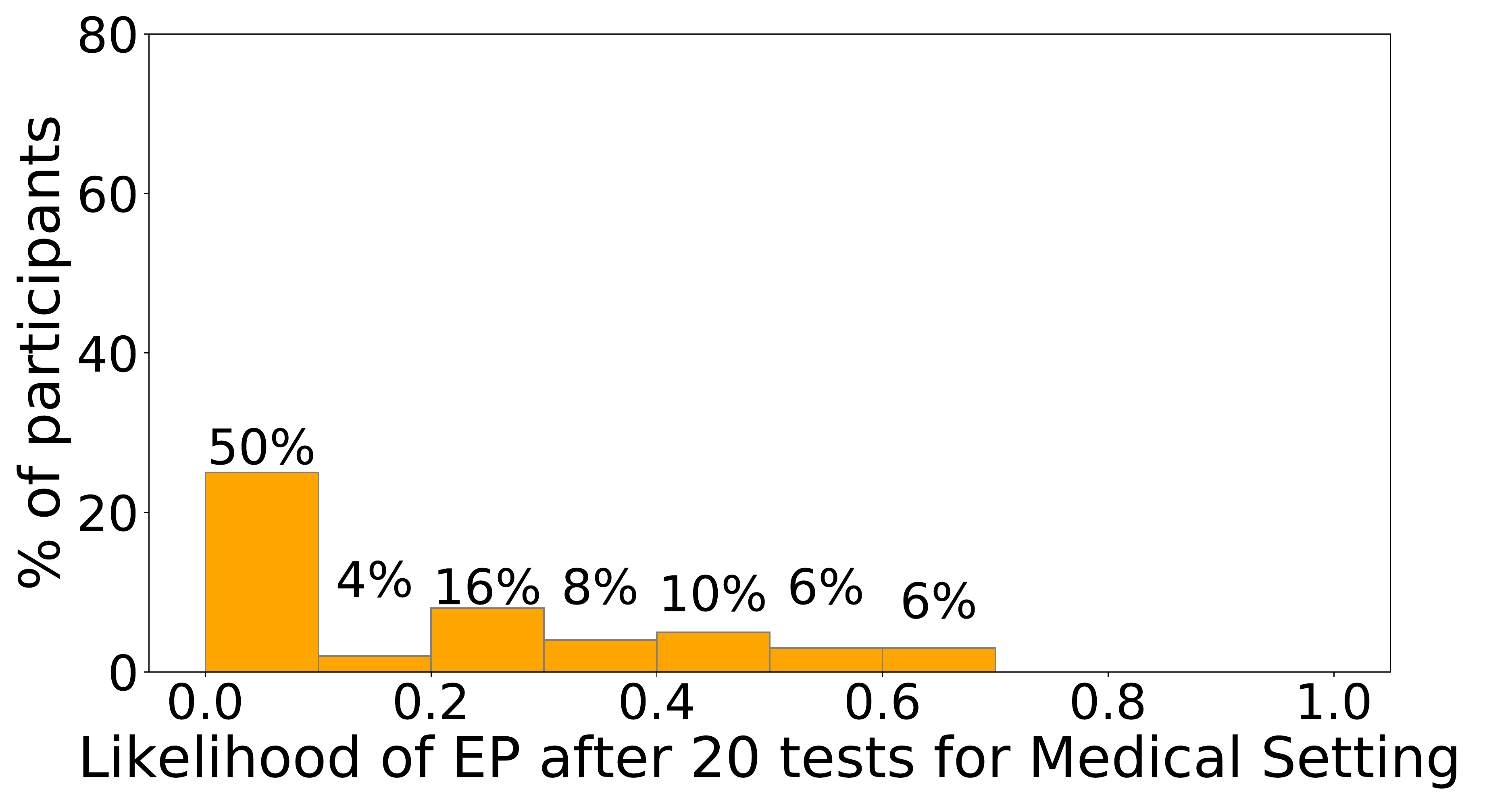}
		\end{subfigure}
		\begin{subfigure}[b]{0.31\textwidth}
			\includegraphics[width=\textwidth]{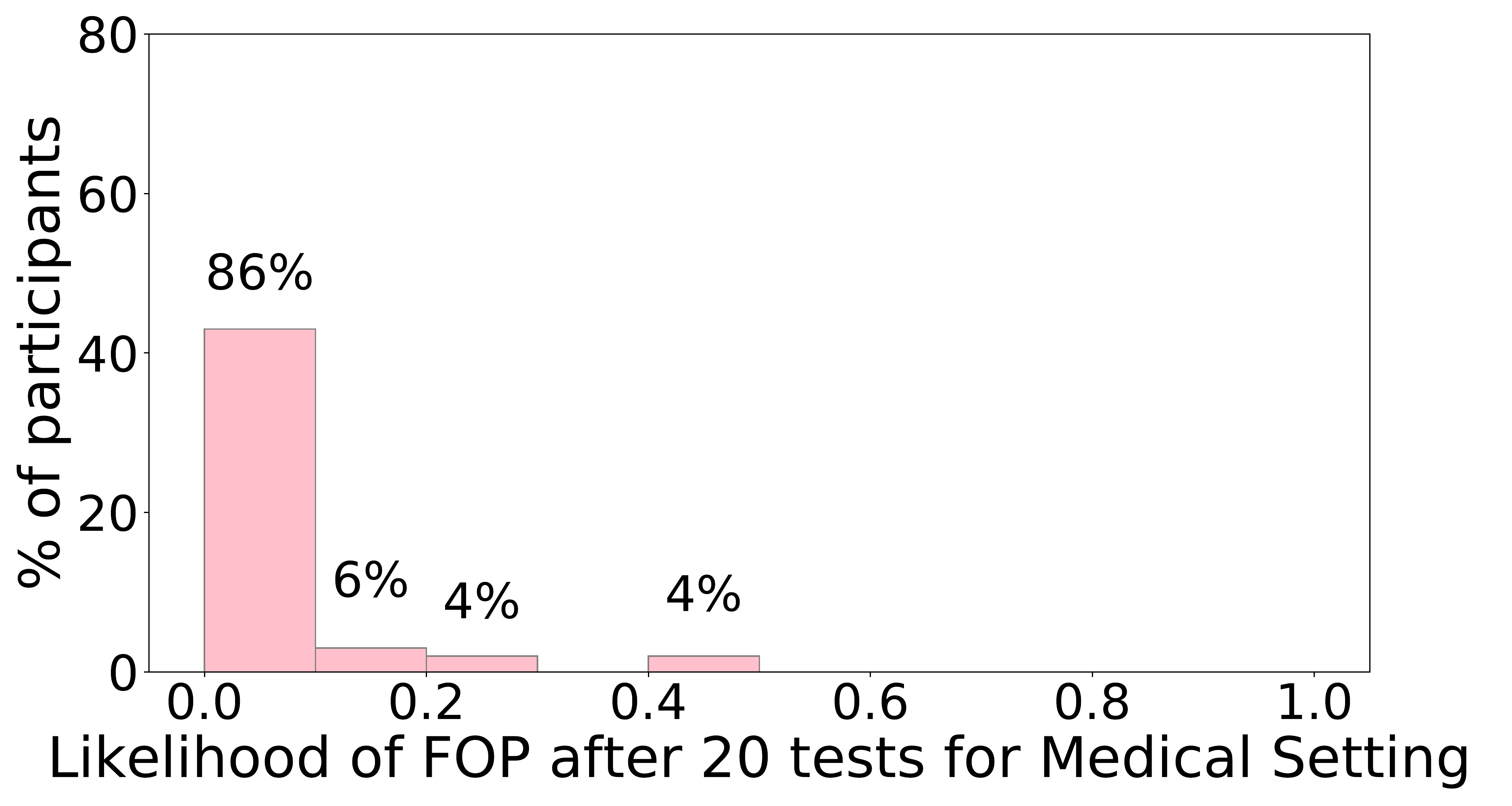}
		\end{subfigure}
		\begin{subfigure}[b]{0.31\textwidth}
			\includegraphics[width=\textwidth]{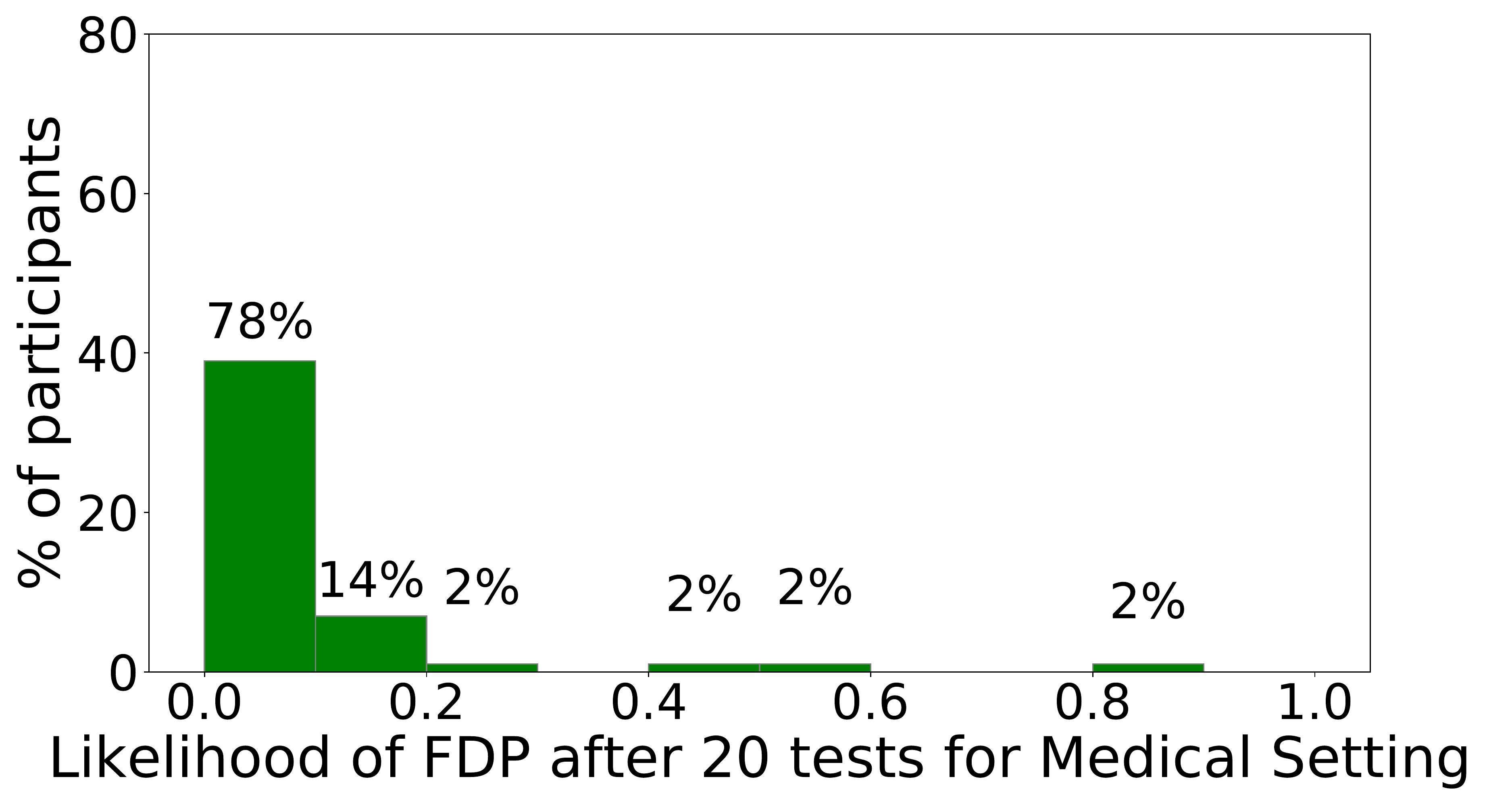}
		\end{subfigure}
		\\
		\begin{subfigure}[b]{0.31\textwidth}
			\includegraphics[width=\textwidth]{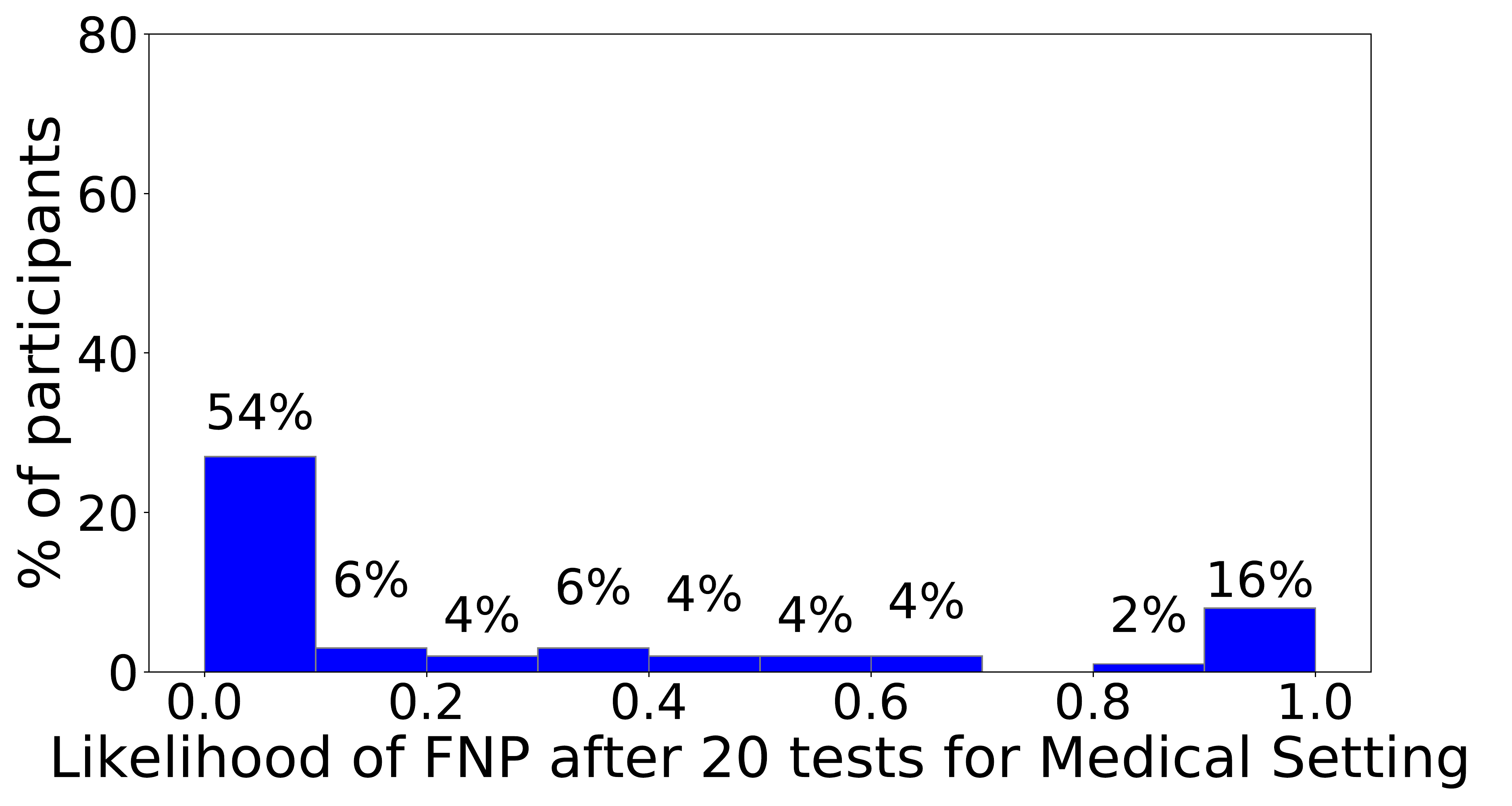}
		\end{subfigure}	  
		\begin{subfigure}[b]{0.31\textwidth}
			\includegraphics[width=\textwidth]{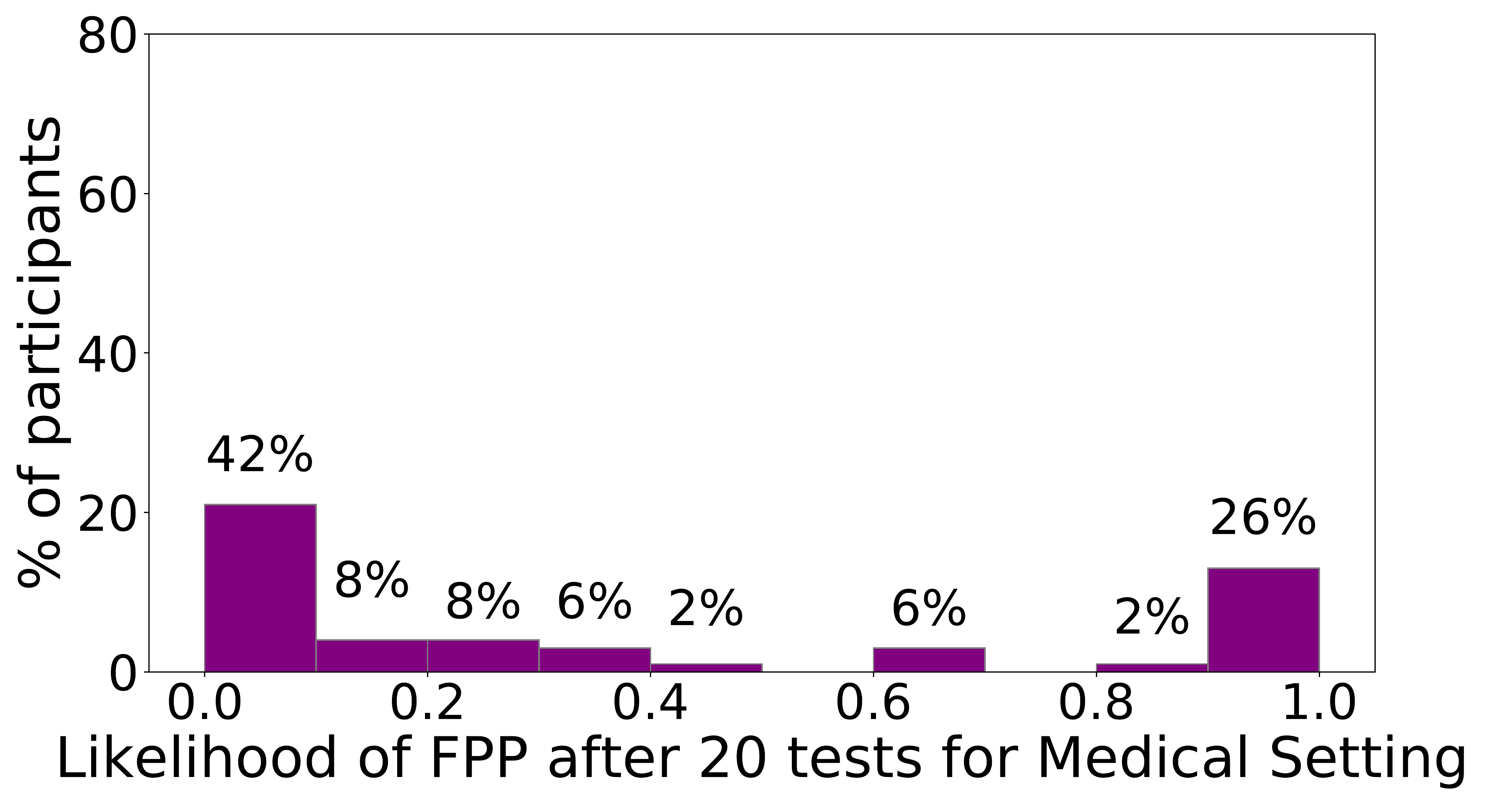}
		\end{subfigure}  
		\caption{The skin cancer risk prediction scenario---the number of participants matched with each notion of fairness (y-axis) along with the likelihood levels (x-axis). False positive parity captures the choices made by the majority of participants. }\label{fig:histogram_med_extra}
	\end{figure*}

\end{document}